\documentclass[oldversion]{aa}
\usepackage{graphicx}
\usepackage{graphics}
\usepackage{txfonts}
\usepackage{natbib}
\bibpunct{(}{)}{;}{a}{}{,} % to follow the A&A style

\begin{document}

   \title{Molecular Gas in NUclei of GAlaxies (NUGA).}

   \subtitle{X. The Seyfert 2 galaxy NGC~3147}

  \author{ V. Casasola\inst{1,2,4}, F. Combes\inst{2}, S. 
Garc\'{\i}a-Burillo\inst{3},  L. K. Hunt\inst{4}, S. L\'eon\inst{5}, \& 
A. J. Baker\inst{6}}

   \offprints{V. Casasola; viviana.casasola@unipd.it}

   \institute{Dipartimento di Astronomia, Universit\`a di Padova, Vicolo 
dell'Osservatorio 2, I-35122, Padova
              %\email{viviana.casasola@unipd.it
\and Observatoire de Paris, LERMA, 61 Av. de l'Observatoire, F-75014 Paris, 
France
\and Observatorio Astron\'omico Nacional (OAN), Alfonso XII, 3, 28014-Madrid, 
Spain
\and Istituto di Radioastronomia/INAF, Largo Enrico Fermi, 5, 50125 Firenze, 
Italy
\and Instituto de Astrof\'{\i}sica de Andaluc\'{\i}a (CSIC), Camino Bajo de 
Hu\'etor, 24, 18008 Granada, Spain
\and Dept. of Physics and Astronomy, Rutgers, the State Univ. of New Jersey, 
136 Frelinghuysen Road, Piscataway, NJ 08854, USA
     }

   \date{Received XXX 2008/ Accepted YYY 2008}
   
   \titlerunning{NGC 3147}

   \authorrunning{Casasola et al.}

% \abstract{}{}{}{}{}
% 5 {} token are mandatory

  \abstract
{We present $^{12}$CO(1-0) and $^{12}$CO(2-1) observations of the SA(rs)bc Seyfert 2 galaxy 
NGC\,3147, obtained with the IRAM interferometer at 1\farcs9\,$\times$1\farcs6 and 
1\farcs6\,$\times$1\farcs4
resolutions, respectively. We have also observed the central region of 
NGC\,3147 with the IRAM 30\,m telescope (at resolutions of 22\arcsec\
and 12\arcsec\ for $^{12}$CO(1-0) and $^{12}$CO(2-1), respectively), in order to 
obtain complete sampling at low spatial frequencies. These observations have 
been made in the context of the NUclei of GAlaxies (NUGA) project, aimed at 
the study of the different mechanisms for gas fueling of active galactic 
nuclei (AGN).  
A central peak seen mainly in $^{12}$CO(2-1) and a ring-like structure
at $r \simeq 10$\arcsec\,$\sim$2\,kpc dominate the $^{12}$CO maps. 
In $^{12}$CO(1-0) an outer spiral at $r \simeq 20$\arcsec\,$\sim$4\,kpc
is also detected, not visible in $^{12}$CO(2-1) emission because it 
falls outside the field-of-view of the primary beam.
The average $I_{21}/I_{10}$ line ratio 
is $\sim0.7$ in temperature units over the region mapped in both lines, 
consistent with the optically thick emission expected in the nuclei of spiral 
galaxies.  The kinematics of the molecular structures are quite regular, although there
is evidence for local non-circular or streaming motions.
We show that the molecular gas distribution is similar 
but not exactly identical to those of star formation tracers, i.e., infrared 
({\it Spitzer}) and ultraviolet ({\it GALEX}) emission. This agreement is 
consistent with a scenario of steady, ongoing formation of stars from the 
molecular clouds at a rate of $\sim 1\,M_\odot\,{\rm yr}^{-1}$ within the
innermost 4\,kpc in radius.\\
Using a near-infrared (NIR) image obtained with adaptive optics at the 
Canada-France-Hawaii Telescope (CFHT), we identify a weak bar in NGC\,3147, 
which is classified as non-barred galaxy in the optical.  We then compute the 
gravity torques exerted by this stellar bar on the gas.  The torque is 
obtained first at each point in the map, and then azimuthally averaged with a
weighting determined by the gas surface density traced by the CO emission.
We find that the gas inside the inner CO ring is subject to a net negative 
torque and loses angular momentum.  This is expected for gas at the 
ultra-harmonic resonance (UHR), just inside the corotation resonance of the 
stellar bar. In contrast, the gas outside corotation, in the spiral arms 
comprising the outer spiral structure, suffers positive torques 
and is driven outwards.
We conclude that some molecular gas is presently flowing into the central region, 
since we find negative torques down to the resolution limit of our
images.

}

   \maketitle

\section{Introduction}

The NUclei of GAlaxies (NUGA) project \citep[][]{garcia03} is an IRAM Plateau 
de Bure Interferometer (PdBI) survey of nearby active galaxies  
to map the distribution and dynamics of molecular gas in the inner 1\,kpc at 
high spatial resolution ($\sim\,0\,\farcs5 - 1^{\prime\prime}$, 
corresponding to $\sim50-100\,{\rm pc}$), and to study the mechanisms for 
gas fueling of low-luminosity active galactic nuclei (AGNs).

Most galaxies possess central supermassive black holes (SMBHs), and gas accretion 
onto these black holes is the phenomenon usually invoked to explain nuclear 
activity in galaxies.  However, even if most galaxies host black 
holes, the existence of nuclear activity is far from universal. It is not 
clear whether the main limiting factor is the global gas mass available for 
fueling the AGN or the mechanisms for efficiently removing the angular 
momentum of the gas.

The main and non-trivial problem linked to the fueling of AGN is the
removal of angular momentum from the disk gas \citep[e.g.,][and
references therein]{jogee06}, a process which
can be accomplished through non-asymmetric perturbations.
These can be perturbations of external origin, such as galaxy collisions,
mergers, and mass accretion \citep{heckman86}, or of internal origin due to 
density waves, such as spirals or bars, and their gravity torques 
\citep[e.g.][]{sakamoto,combes01}.
In addition to primary bars, fueling processes can be associated with more 
localized phenomena, such as nested nuclear bars \citep[e.g.][]{friedli}, 
lopsidedness or $m = 1$ perturbations \citep[e.g.][]{shu,garcia00}, or warped 
nuclear disks \citep[e.g.][]{schinnerer00a,schinnerer00b}.

Since molecular gas is the predominant phase of the interstellar medium (ISM) 
in the inner kiloparsec of spiral galaxies, the study of its morphology and 
dynamics represents an optimal possibility for investigating AGN fueling 
mechanisms and their link with circumnuclear star formation.  In order to 
do this, high-resolution maps of molecular gas are needed.  Previous 
CO single dish \citep[][]{heckman89,young,braine93,casoli96,vila98}
and interferometric \citep[][]{sakamoto,regan01,helfer03} surveys have mapped 
the gas in galaxies with relatively low spatial resolution ($4^{\prime\prime} 
-7^{\prime\prime}$). Moreover, the majority of these surveys only included 
small numbers of AGN in their samples. This paper, dedicated to the galaxy 
NGC\,3147, is the tenth of a series where results obtained for the galaxies 
in the NUGA sample are described on a case-by-case basis.

 \begin{table}
   \caption[]{Fundamental parameters for NGC\,3147.}
   \begin{center}
   \begin{tabular}{lll}
   \hline
   \hline
     Parameter  & Value & References$^{\mathrm{a}}$ \\
   \hline
   RA (J2000) & 10$^h$16$^m$53.6$^s$ & (1) \\
   DEC (J2000) & 73$^{\circ}$24$^{\prime}$03$^{\prime\prime}$ & (1) \\
   $V_{\rm hel}$ & 2813 km\,s$^{-1}$ & (1) \\
   RC3 Type & SA(rs)bc & (1) \\
   T Hubble Type & 3.88 & (2) \\
   Inclination & 29.5$^{\circ}$ & (2) \\
   Position Angle & 150$^{\circ}$ & (2) \\
   Distance & 40.9\,Mpc ($1^{\prime\prime} = 198\,{\rm pc}$) & (1) \\
   $L_{B}$ & $4.3 \times 10^{10}\,L_{\odot}$ & (3) \\
   $M_{\rm H\,I}$ & $8.3 \times 10^{8}\,M_{\odot}$ & (3) \\
   $M_{\rm dust}$(60 and 100\,$\mu$m)& $1.7 \times 10^{6}\,M_{\odot}$ & (3) \\
   $L_{\rm FIR}$ & $3.4 \times 10^{10}\,L_{\odot}$ & (4) \\
   \hline
   \hline
   \end{tabular}
   \label{table1}
   \end{center}
   \begin{list}{}{}
   \item[$^{\mathrm{a}}$] (1) NASA/IPAC Extragalactic Database (NED); 
   (2) Lyon Extragalactic Database (LEDA); (3) \citet{bettoni}; (4) {\it IRAS} Catalog.
   \end{list} 
   \end{table}
 
NGC\,3147 ($D = 40.9\,{\rm Mpc}$, $1^{\prime\prime} = 198\,{\rm pc}$, 
$H_0 = 75\,{\rm km\,s^{-1}\,Mpc}^{-1}$) 
is an isolated \citep{bettoni} Seyfert 2 galaxy \citep{ho} of Hubble 
morphological type SA(rs)bc.  Both \textit{ROSAT} \citep{roberts} and 
\textit{Chandra} \citep{terashima03} observations have shown that this galaxy 
possesses a pointlike nuclear X-ray source,
which is presumed to be the location of the 
AGN. The 0.3--10\,keV \textit{Chandra} image reveals a 
bright, compact source surrounded by very faint, soft, and diffuse emission. The 
2--10\,keV core is clearly detected and confined to a region of
$2^{\prime\prime}$ in radius.  
Interferometric observations of the continuum emission at cm wavelengths 
with MERLIN and the VLBA have also shown a pointlike nonthermal continuum 
source at the position of the nucleus of NGC\,3147,
coincident with the X-ray source \citep[][]{ulvestad,krips06,krips07a}.
Very recent optical and X-ray observations of the 
nuclear region of this galaxy suggest 
that NGC\,3147 is the first ``true'' Seyfert 2 in the sense that it 
intrinsically lacks a broad-line region \citep{bianchi08}. 
Table \ref{table1} reports the fundamental characteristics of
NGC\,3147.

The structure of this paper is as follows.  In Sect. 2, we describe our new
observations of NGC\,3147 and the literature data with which we compare them.
In Sects. 3 and 4, we present the observational results, both single dish
and interferometric, describing the principal properties of the molecular gas
including its morphology, its kinematics, and its
excitation. A comparison between the CO observations and those obtained at
other wavelengths is given in Sect. 5.  In Sect. 6, we describe the
computation of the gravity torques exerted on the gas by a weak stellar bar
that we have identified using a near-infrared (NIR) image. In Sect. 7, we discuss
our principal observational and theoretical results, which are summarized
in Sect. 8.
 
 \begin{figure*}
	\centering
        \includegraphics[width=0.9\textwidth,angle=-90]{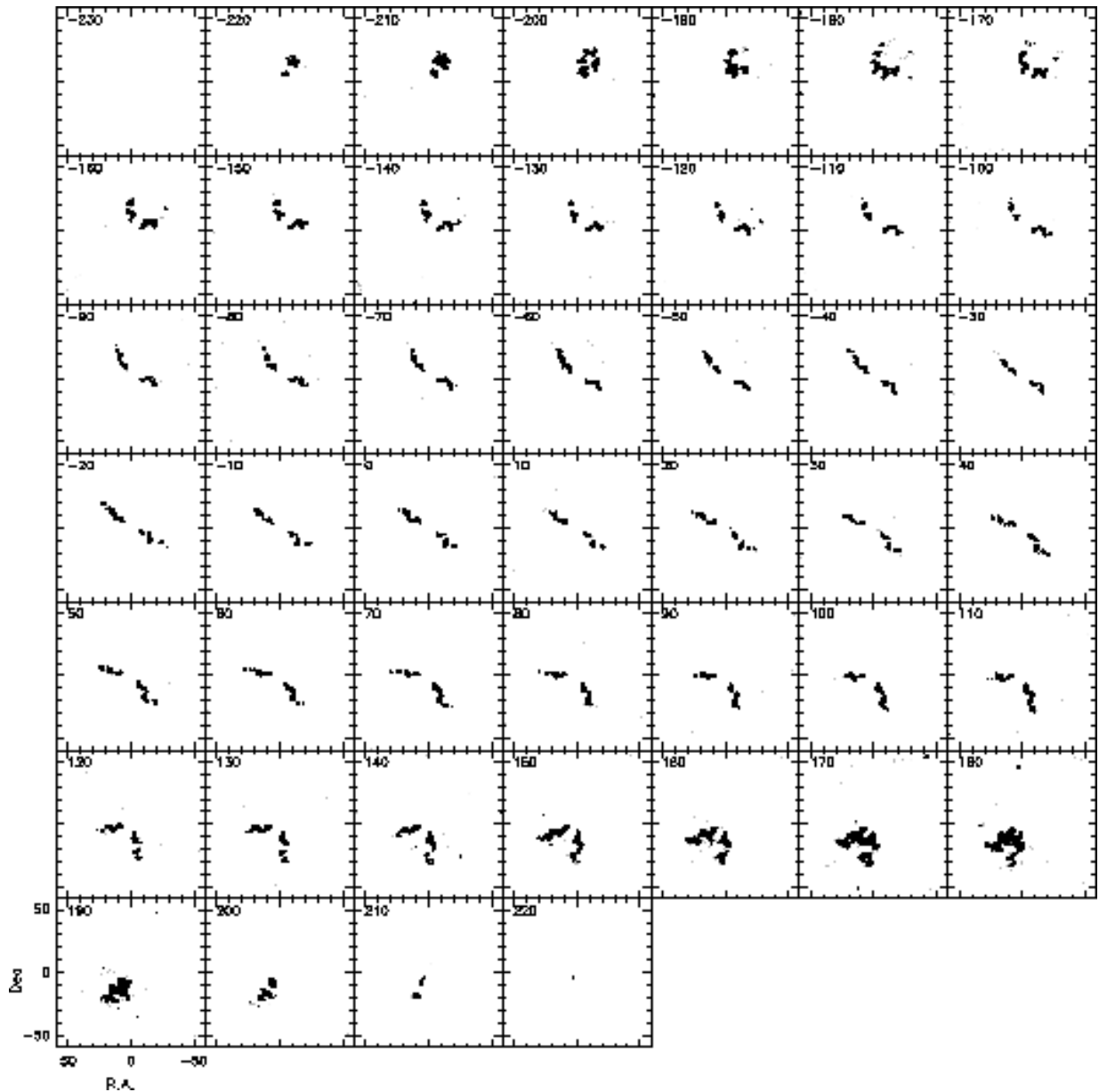}
	\caption{$^{12}$CO(1-0) velocity channel maps observed with the IRAM PdBI+30\,m
in the nucleus of NGC\,3147, with a spatial resolution of 1\farcs9
$\times$ 1\farcs6 (HPBW). The center of the maps, given in Table
\ref{table1}, is $\alpha_{2000}$ = 10$^h$16$^m$53.6$^s$, $\delta_{2000}$ =
73$^{\circ}$24$^{\prime}$03$^{\prime\prime}$.  Velocity channels range from
$\Delta V = -230\,{\rm km\,s^{-1}}$ to $+220\,{\rm km\,s^{-1}}$ in steps of
$10\,{\rm km\,s^{-1}}$ relative to $V_{\rm hel} = 2813\,{\rm km\,s^{-1}}$.
The contours begin at $5\,{\rm mJy\,beam^{-1}}$, their spacing is $5\,{\rm
mJy\,beam^{-1}}$, and their maximum is $50\,{\rm mJy\,beam^{-1}}$.}
	\label{channel}
  \end{figure*}

\begin{figure*}
	\centering
        \includegraphics[width=0.9\textwidth,angle=-90]{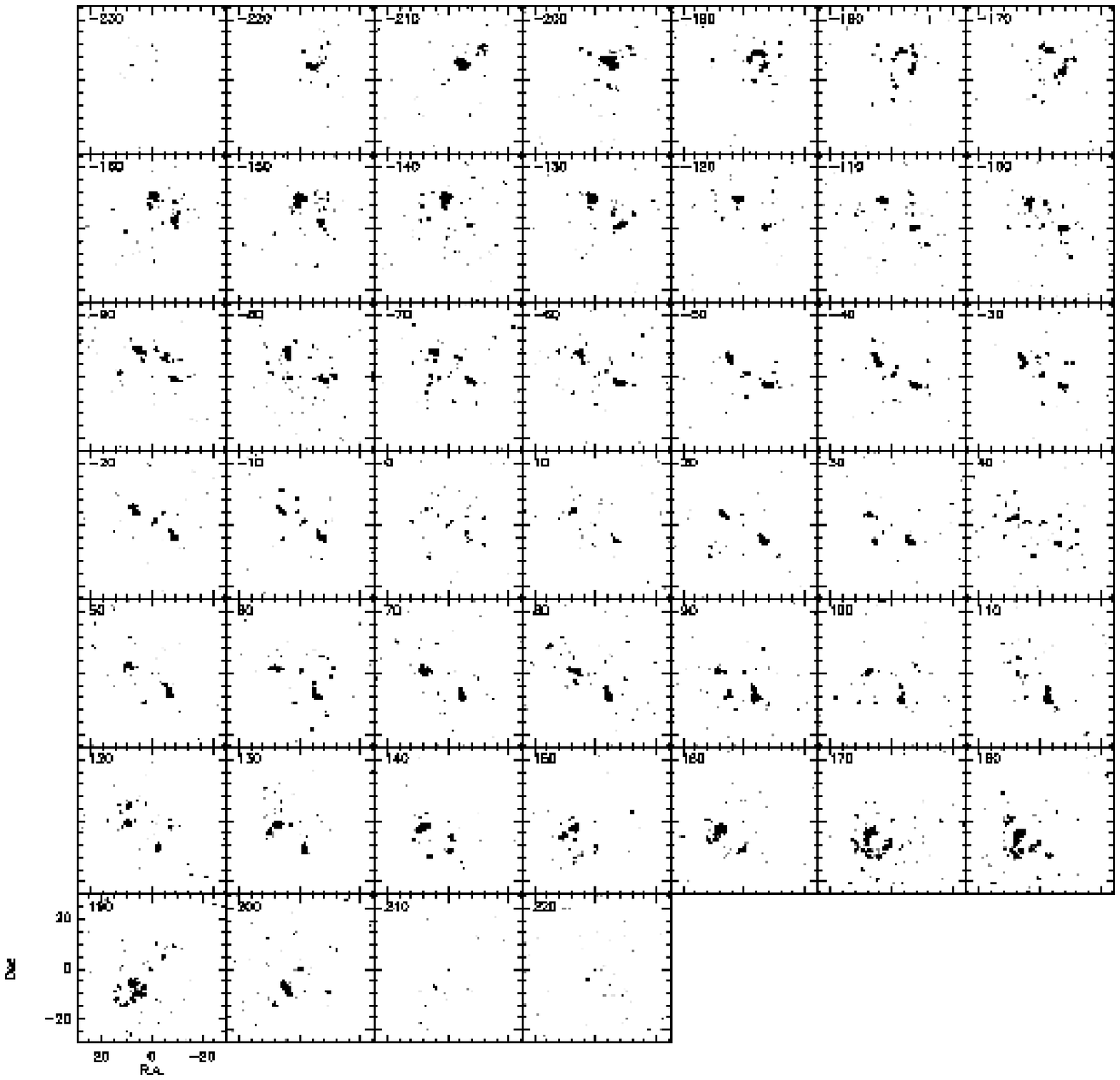}
	\caption{Same as Fig. \ref{channel} but for the $^{12}$CO(2-1) line, 
        with a spatial resolution of 1\farcs6 $\times$ 1\farcs4.
        The velocity range is the same as for Fig. \ref{channel}.
        The contours begin at $10\,{\rm mJy\,beam^{-1}}$, their spacing is $5\,{\rm
        mJy\,beam^{-1}}$, and their maximum is $45\,{\rm mJy\,beam^{-1}}$.}
	\label{channel21}
         \end{figure*}

 \section{Observations}

 \subsection{IRAM interferometric CO observations}
 
We observed the $J = 1-0$ and $J = 2-1$ lines of ${}^{12}$CO in NGC\,3147
using the IRAM PdBI in October 2004, with the array deployed in its ABCD
configurations.  The six 15\,m antennae were equipped with dual-band SIS
receivers yielding SSB receiver temperatures between 40 and 50\,K at both
frequencies. The precipitable water vapor ranged from 4 to 10\,mm (i.e.,
giving opacities of $\sim 0.2-0.3$), resulting in system temperatures of
approximately $200-300\,{\rm K}$ on average. The spectral correlators were
centered at 114.197\,GHz ($2793.75\,{\rm km\,s^{-1}}$) and 228.390\,GHz
($2793.27\,{\rm km\,s^{-1}}$), respectively.  The coordinates of the PdBI
phase tracking center are given in Table \ref{table1} and correspond to the
(6\,cm VLBA) nuclear radio position  of \citet{nagar02}.

\begin{figure}[h]
 \centering
 \includegraphics[width=\columnwidth,angle=0,bb=47 145 590 680]{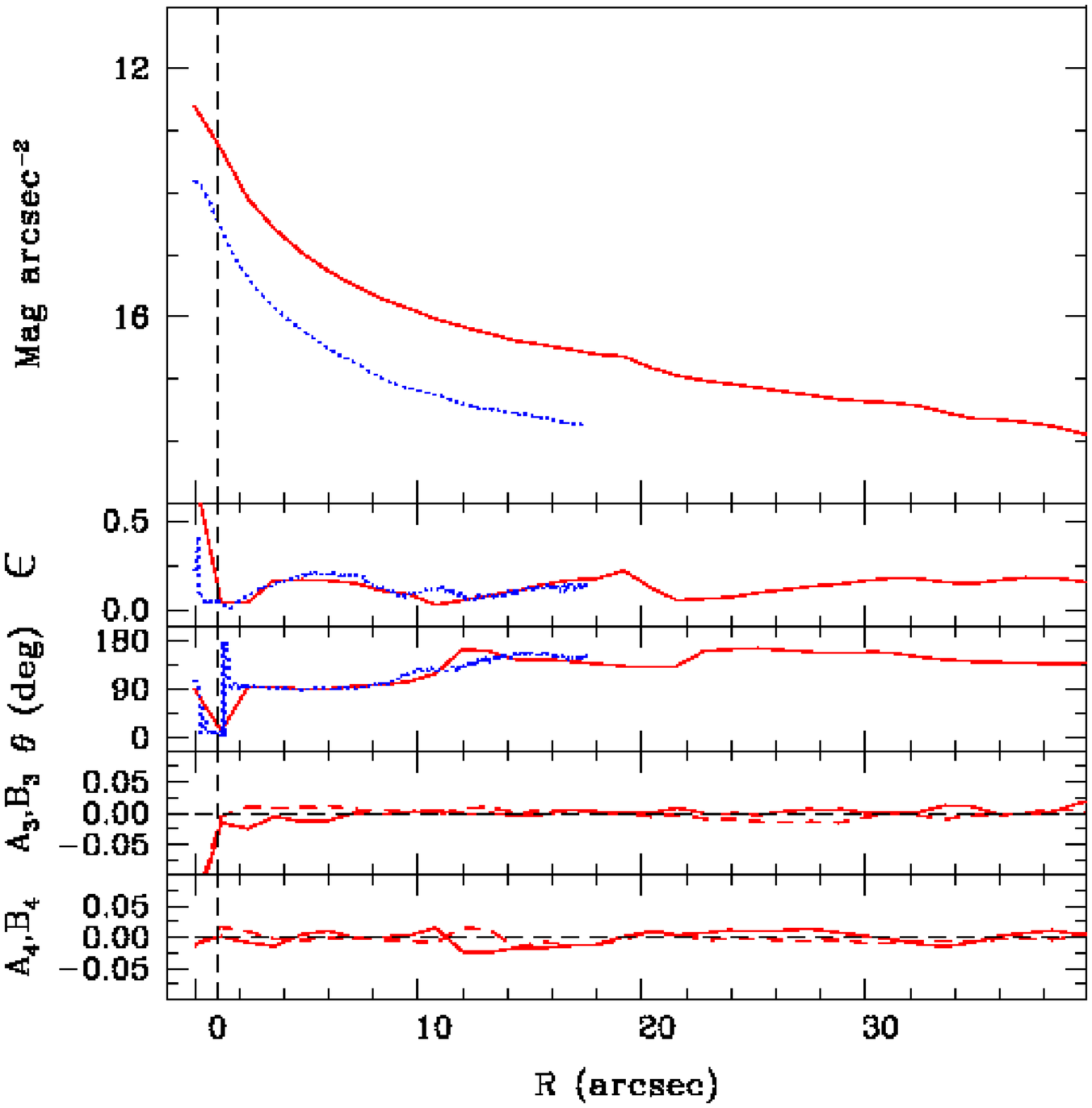}
 \caption{Radial brightness profiles are shown in the
 top panel, with $J$ given as a
 dotted line (blue), and IRAC $3.6\,{\rm \mu}m$ as a solid
 one (red).
 The vertical dashed line corresponds to 200\,pc.
 The lower panels display the radial runs of ellipticity, $\epsilon$,
 position angle $\theta$, and cos($4\theta$) terms in the ellipse
 fitting residuals.
 \label{fig:ellipse} 
 }
 \end{figure} 

Data cubes with 256 $\times$ 256 pixels
(0\,\farcs46\,${\rm pixel}^{-1}$ for $^{12}$CO(1-0) and
0\,\farcs23\,${\rm pixel}^{-1}$ for $^{12}$CO(2-1)) were created 
over a velocity interval of -240 km\,s$^{-1}$ to +240 km\,s$^{-1}$
in bins of 10 km\,s$^{-1}$.
The images were reconstructed
using the standard IRAM/GILDAS\footnote{http://www.iram.fr/IRAMFR/GILDAS/} 
software \citep{guilloteau} and restored with Gaussian beams of dimensions 
1\farcs8 $\times$ 1\farcs6 (PA = $62^{\circ}$) at 115\,GHz 
and 1\farcs4 $\times$ 1\farcs2 (PA = $61^{\circ}$) at 
230\,GHz. In the cleaned maps, the rms levels are $2\,{\rm mJy\,beam^{-1}}$ 
and $4\,{\rm mJy\,beam^{-1}}$ for the $^{12}$CO(1-0) and $^{12}$CO(2-1) lines, respectively 
(at a velocity resolution of $10\,{\rm km\,s^{-1}}$).  The conversion 
factors between intensity and brightness temperature are $32\,{\rm 
K\,(Jy\,beam^{-1})^{-1}}$ at 115\,GHz and $14\,{\rm K\,(Jy\,beam^{-1})^{-1}}$ 
at 230\,GHz.

A continuum point source is detected at both 3\,mm and 1.3\,mm.
The continuum flux density is $5.2 \pm 0.2\,{\rm mJy}$ at 3\,mm
and $2.2 \pm 0.4\,{\rm mJy}$ at 1.3\,mm. Both values
are consistent with the synchrotron source detected at centimeter
wavelengths for a power-law spectrum with a slope of $-1$.
All maps presented in the paper are continuum-subtracted.

 \subsection{IRAM single dish CO and HCN observations}
 \label{singleo}

 We performed IRAM 30\,m telescope observations in a $5 \times 5$
 raster pattern with $7^{\prime\prime}$ spacing in 
 July 2002 and in June 2004. We used 4 SIS receivers to observe simultaneously 
 at the frequencies of the $^{12}$CO(1-0) (115\,GHz), the $^{12}$CO(2-1) (230\,GHz), and the 
 HCN(1-0) (89\,GHz) lines. The half power beam widths are
 $22^{\prime\prime}$ for $^{12}$CO(1-0), $12^{\prime\prime}$ for $^{12}$CO(2-1), and
 $29^{\prime\prime}$ for HCN(1-0). The typical system temperatures
 were $\sim250 \,{\rm K}$ at 115\,GHz, $\sim350-700\,{\rm K}$ at
 230\,GHz, and $\sim120$\,K at 89\,GHz.  The line intensity scale throughout
 this paper is expressed in units of $T_{\rm mb}$, the beam-averaged radiation
 temperature. The value of $T_{\rm mb}$ is related to $T^{*}_{A}$, the
 equivalent antenna temperature (corrected  for rear spillover and ohmic
 losses) reported above the atmosphere, by $\eta=T^{*}_{A}/T_{\rm mb}$ where
 $\eta$ is the telescope main-beam efficiency.  $\eta$ is 0.79 at 115\,GHz,
 0.57 at 230\,GHz, and 0.82 at 89\,GHz.  All observations were performed in
 ``wobbler-switching'' mode, with a minimum phase time for spectral line
 observations of 2\,s and a maximum beam throw of $240^{\prime\prime}$. The
 pointing accuracy was $\sim3^{\prime\prime}$ rms.

 Single dish observations were used to compute short-spacings and
 complete the interferometric measurements \citep[e.g.][]{combes04}. In
 particular, short-spacing visibilities are computed from a map built by
 interpolation of the 30\,m beam and multiplied by the PdBI primary beam.
 We combined 30\,m and PdBI data, using the SHORT-SPACE task in the GILDAS
 software. To find the best compromise between good angular resolution and
 complete restoration of the missing extended flux, we varied the weights
 attached to the 30\,m and PdBI data.
 After writing the combined datasets to visibility tables, 
 converting to maps using standard data reduction procedures, 
 and deconvolving using the Clark algorithm, we obtained maps with 
 angular resolutions of 1\farcs88 $\times$ 1\farcs63 
 at PA 61.5$^{\circ}$ for the $^{12}$CO(1-0) map and 
 1\farcs61 $\times$ 1\farcs4 at PA 65.4$^{\circ}$ for
 the $^{12}$CO(2-1) map.
 The weights were adjusted in order to obtain the same mean
 weights in the single-dish data as in the interferometric data in the $uv$
 range of $1.25\,D/\lambda$ to $2.5\,D/\lambda$ ($D = 15$\,m). All figures
 presented in this paper are made with short-spacing-corrected data.

Figs. \ref{channel} and \ref{channel21} display the channel maps 
for the $^{12}$CO(1-0) and $^{12}$CO(2-1) lines, respectively.
All maps are centered on the position of Table \ref{table1}, and the
dynamical center coincides with this center, which is also the position 
of the AGN (the radio continuum source).

\begin{figure*}
 \centering
 \includegraphics[height=0.75\textwidth,angle=-90]{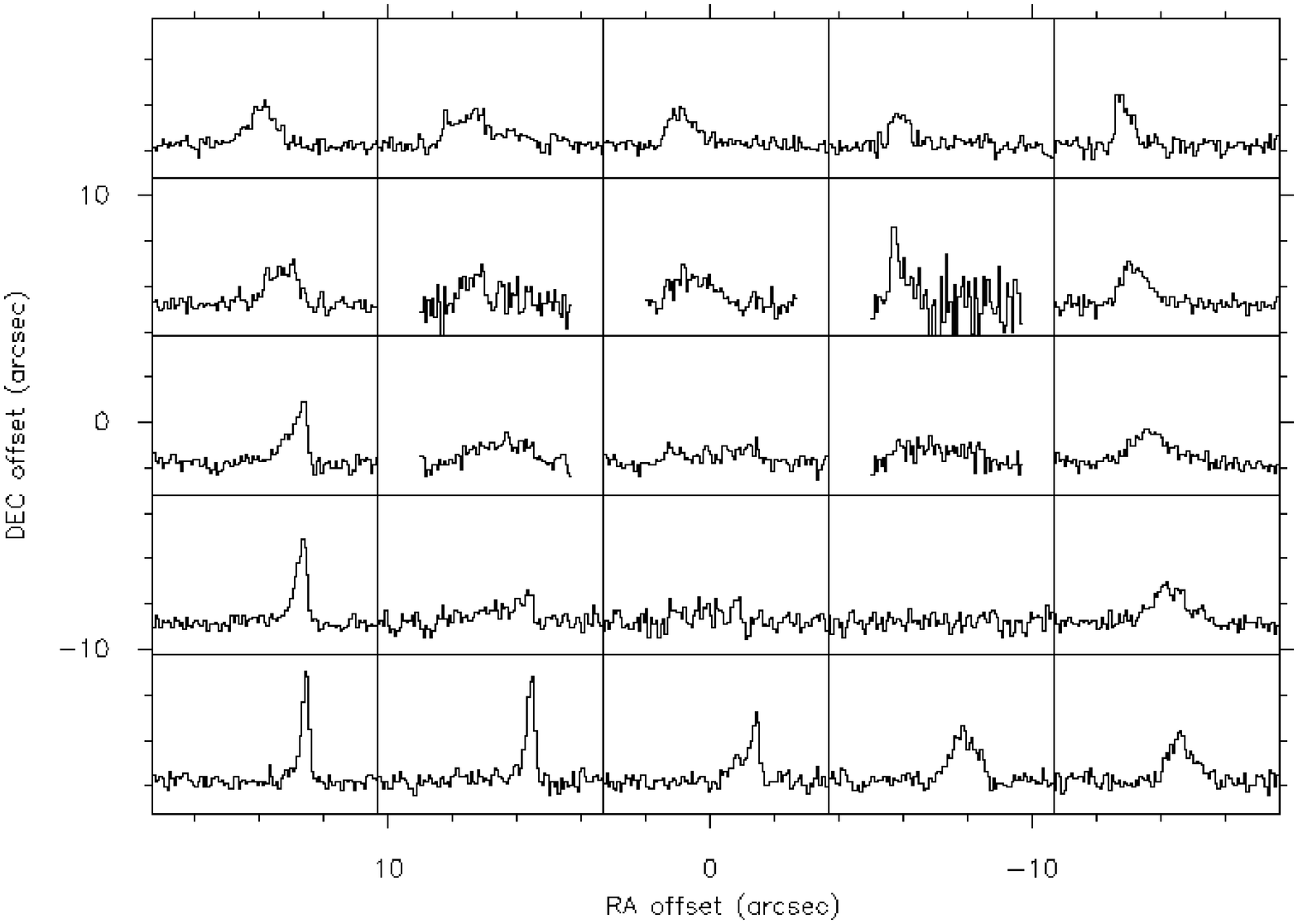}
 \caption{Spectra maps of NGC\,3147 made with the IRAM 30\,m with
 $7^{\prime\prime}$ spacing in $^{12}$CO(1-0) (top) and $^{12}$CO(2-1) (bottom). The
 positions are offsets relative to the center of NGC\,3147, whose coordinates
 are listed in Table \ref{table1}. Each spectrum has a velocity scale from
 $-500$ to $500\,{\rm km\,s^{-1}}$, and an antenna temperature scale
 ($T^{*}_{A}$) from $-0.05$ to $0.25\,{\rm K}$ for $^{12}$CO(1-0) and to 0.20\,K
 for $^{12}$CO(2-1).}
 \label{co21-30m}
 \end{figure*}

Figs. \ref{co21-30m} and \ref{n3147-hcn} display the single dish data,
for the $^{12}$CO(1-0), $^{12}$CO(2-1), and HCN(1-0) lines. The two CO lines have been 
mapped on a $5 \times 5$ grid with $7^{\prime\prime}$ spacings, while 
the HCN(1-0) line has been mapped on a $3 \times 3$ grid with 
$7^{\prime\prime}$ spacings and the nine HCN(1-0) spectra have been 
averaged to improve the signal-to-noise.

 To estimate the flux filtered out by the interferometric observations, we
 have computed the flux measured by \citet{young} with the FCRAO 14\,m
 telescope (HPBW = $45^{\prime\prime}$).  They found a $^{12}$CO(1-0) intensity
 towards the center of $I({\rm CO}) = 5.75\,{\rm K\,km\,s^{-1}}$ (in the
 $T^{*}_{A}$ scale), which corresponds to an integrated flux of $S({\rm CO}) =
 242\,{\rm Jy\,km\,s^{-1}}$ if we adopt a conversion factor of $42\,{\rm
 Jy\,K^{-1}}$. 
 In the same region, we measure $130\,{\rm
 Jy\,km\,s^{-1}}$ in $^{12}$CO(1-0) with the PdBI alone, and $290 \,{\rm
 Jy\,km\,s^{-1}}$  with the IRAM 30\,m in a beam of $22^{\prime\prime}$.
 With short-spacings added, we estimate the total flux to be
 $295\,{\rm Jy\,km\,s^{-1}}$ in $^{12}$CO(1-0), which is compatible with the 
 FCRAO value within the uncertainties.
 In $^{12}$CO(2-1), the PdBI alone indicates 
 $30\,{\rm Jy\,km\,s^{-1}}$ with a large region with negative flux 
 levels. With the IRAM 30\,m in $^{12}$CO(2-1) we measure a total 
 flux of $140\,{\rm Jy\,km\,s^{-1}}$ in a beam of 12$^{\prime\prime}$, 
 and when short-spacings are included we recover the value of 
 $151\,{\rm Jy\,km\,s^{-1}}$.
 
 The implication is that much of the 
 $^{12}$CO(2--1) emission comes from smoothly distributed gas.

 \begin{figure*}
 \centering
 \includegraphics[height=0.75\textwidth,angle=-90]{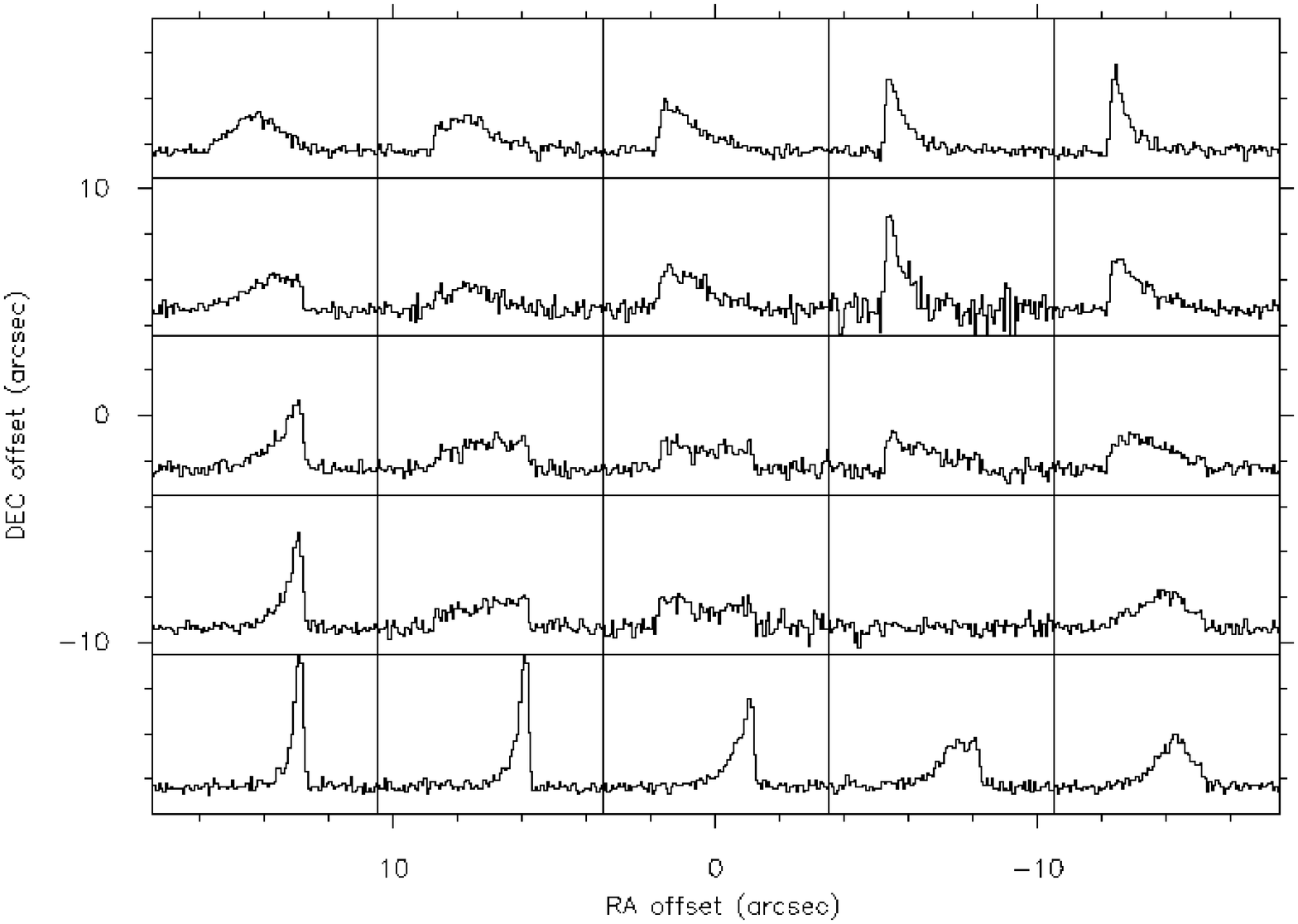}
 \label{co10-30m}
 \end{figure*}

 \subsection{Near-infrared observations}

For modeling the gravitational potential, we used NIR images 
which have the advantage of being relatively free of dust extinction and more
representative of old stellar populations than blue or visible images. 

The NIR images were obtained in December 1998 at the 3.6\,meter
Canada-France-Hawaii Telescope (CFHT), using the CFHT Adaptive Optics
Bonnette (AOB) and the KIR infrared camera.
The AOB, also called PUEO after the sharp-visioned Hawaiian owl,
delivers essentially diffraction-limited images at $H$ and $K'$
(0\,\farcs11 and 0\,\farcs14, respectively), and images 
with FWHM $\sim0\,\farcs1$ at $J$ with guide stars as faint as $R=14$ 
\citep[see][]{rigaut98}. Here the Seyfert nucleus of the galaxy 
was used for wavefront sensing.
The KIR 1024 $\times$ 1024 pixel HgCdTe array 
has 0\,\farcs035/pixel, providing a field of view of 
$36^{\prime\prime} \times 36^{\prime\prime}$.
The observations were obtained in excellent seeing conditions 
($\sim0\,\farcs6$ in the $V$ band), with several images 
taken in a dithering procedure to correct for camera defects. 
The total on-source integration time was 8 minutes 
each for $J$ and $K^{\prime}$, and the total observing time for 
the two filters was 50 minutes including sky measurements and overheads.

A $J-K$ map was constructed by first subtracting sky
emission, as estimated from the outer regions of the images.
Then mean instrumental magnitude zero points were used to calculate a
$J-K$ image, after aligning them to a common center.

\subsection{Archival observations with \textit{Spitzer} and \textit{GALEX}\label{sec:archive}}

We acquired public images of NGC\,3147 at other wavelengths to compare our
molecular gas observations with other star formation tracers.
 
We used a far-ultraviolet (FUV) image from the \textit{GALEX} satellite, 
whose band is centered at $\lambda_{\rm eff} = 1516\,{\rm \AA}$.
This image has been already used and studied in the context of 
the \textit{GALEX} Nearby Galaxies Survey 
\citep[NGS,][]{gildepaz04,bianchi03a,bianchi03b}, a project that spans a 
large range of physical properties and morphological types of  
nearby galaxies.
The image has been obtained with a total exposure 
time of 1693\,s and covers a square 
region on the sky of size $\sim 5500^{\prime\prime} \times 
5500^{\prime\prime}$, i.e., much larger than the extent of the optical disk 
of NGC\,3147, with 1\farcs5 pixels.
As the image was reduced with the \textit{GALEX} data pipeline,
it is already expressed in intensity units and sky-subtracted.  
The total FUV 
calibrated magnitude is $14.99 \pm 0.01$, corresponding to a FUV  
flux density of $3669 \pm 17\,{\rm \mu Jy}$.

We also acquired 
infrared (IR) images obtained with the IRAC camera on \textit{Spitzer}, 
available thanks to the project `A Mid-IR Hubble Atlas of Galaxies' 
\citep[Principal Investigator: G. Fazio, see also][]{pahre04}.
These images range in wavelength from 
$3.6\,{\rm \mu m}$ to $8\,{\rm \mu m}$, and were 
reduced with the \textit{Spitzer} data pipeline 
(version S14.0.0). As in the case of the {\it GALEX} image, 
they cover a large sky area ($\sim1400^{\prime\prime} \times 
700^{\prime\prime}$). 
These images have been rotated to canonical 
orientation (north  up, east left) in order to allow correct superposition 
on the molecular gas map.
The pixel size is 1\farcs20 in the final images.
The stellar component at 3.6\,$\mu$m and 4.5\,$\mu$m was 
subtracted from the 8.0\,$\mu$m image 
according to the prescriptions of
\citet{helou04} and \citet{pahre04}.
This procedure provides a ``dust-only'' image at
8\,$\mu$m, which we will discuss in Sect. \ref{sec:ring}.

\begin{figure}[!h]
 \centering
 \includegraphics[height=\columnwidth,angle=-90]{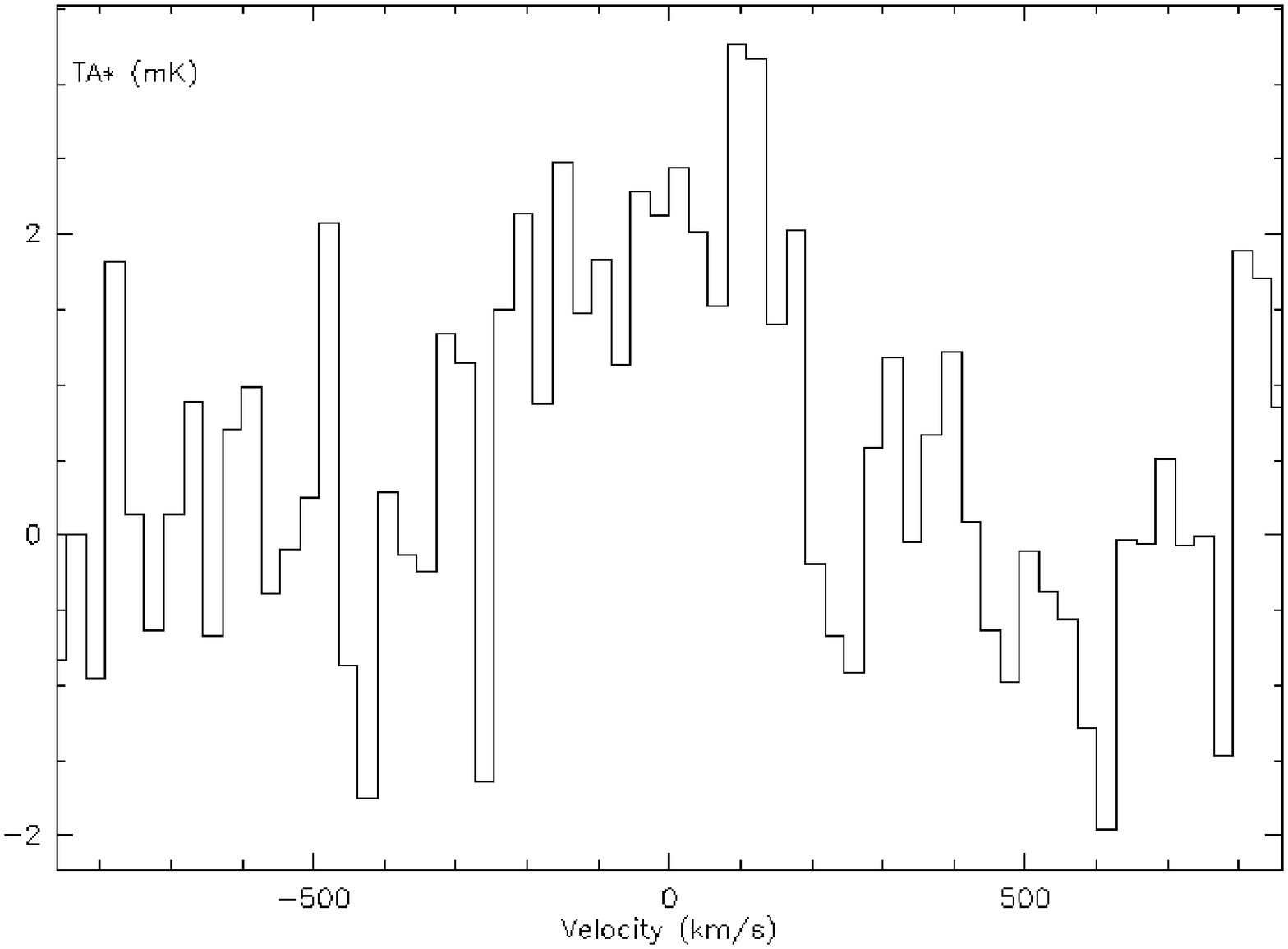}
 \caption{HCN(1-0) spectrum towards the center of NGC\,3147, averaged over
the 9-point map made with the IRAM 30\,m with $7^{\prime\prime}$ spacing.}
 \label{n3147-hcn}
 \end{figure}

Using the $3.6\,{\rm \mu m}$ image,
we calculated radial surface-brightness profiles by fitting
ellipses (using the IRAF/STSDAS task \textit{ellipse}).
The center was fixed, but the ellipticity and the position angle
were allowed to vary.
This brightness profile is shown in Fig. \ref{fig:ellipse}, together with
the $J$-band radial profile obtained in a similar way from the
CFHT image.
The zero point for the IRAC image was taken from the
IRAC documentation at the \textit{Spitzer} Science Center 
(http://ssc.spitzer.caltech.edu/irac/calib/).

No corrections were made to the ``standard'' IRAC photometric calibrations.

Also shown are the runs of ellipticity, ellipse position angle,
and cos($4\theta$) residuals of the ellipse fitting.
A local peak in the ellipticity profile at R\,$\simeq\,5^{\prime\prime}$
(1\,kpc) at constant position angle
(85$^\circ$)
corresponds to a weak bar/oval feature that will be discussed
in Sect. \ref{sec:bar}.

 \section{Single dish results}
 \label{singler}
 The observations 
 performed with the A and B receivers of the IRAM 30\,m telescope in the two $^{12}$CO lines covered the
 central $\sim$50$^{\prime\prime}$, corresponding to the central $\sim$10 kpc
 (in diameter) of the galaxy (Fig. \ref{co21-30m}).
 The 25 observed positions show that the central region of NGC~3147 presents extended molecular 
 emission in both $^{12}$CO(1-0) and $^{12}$CO(2-1) (Fig. \ref{co21-30m}). The maximum 
 detected $T_{mb}$ is 0.32 K in $^{12}$CO(1-0) and 0.31 K in $^{12}$CO(2-1), in the southeast region
 corresponding to the offsets (10$^{\prime\prime}$,-10$^{\prime\prime}$) 
 and (5$^{\prime\prime}$,-10$^{\prime\prime}$) relative to the galaxy center.

 The combination of observations of the $^{12}$CO(1-0) and $^{12}$CO(2-1) lines 
 allowed us to compute the line ratio, 
 $R_{21}=I_{21}/I_{10}=\int{T_{mb}(2-1)}$dv /$\int{T_{mb}(1-0)}$dv, 
 an indicator of physical properties of the molecular gas such as 
 excitation and optical depth.
 After smoothing the $^{12}$CO(2-1) data to the $^{12}$CO(1-0) beam resolution 
 ($22^{\prime\prime}$) and correcting for the different beam efficiencies, we found a mean 
 line ratio for the central region of NGC\,3147 of  0.8.
 This value is consistent with the optically thick emission from molecular 
 gas expected for the central regions of spiral galaxies \citep{braine92}.

 Assuming a CO-H$_{2}$ conversion factor $X = N(H_{2})/I_{CO} = 2.2 \times 10^{20}$ cm$^{-2}$ 
 (K km s$^{-1}$)$^{-1}$ \citep{solomon}, the total H$_{2}$ mass derived from
 these
 single dish observations is 
 M$_{H_{2}}$ = 8.2 $\times$ 10$^9$ M$_{\odot}$, within a radius of 5 kpc.
 
 The HCN(1-0) line has been observed for 9 positions with 7$^{\prime\prime}$ spacing
 covering the central 43$^{\prime\prime}$, corresponding to the central $\sim$8.5 kpc
 (in diameter). The HCN(1-0) emission has been detected most significantly
 in the northwest part of the observed grid 
 corresponding to the offset (-7$^{\prime\prime}$,7$^{\prime\prime}$).
 The HCN(1-0) average spectrum over the 3$\times$3 grid is plotted in Fig. \ref{n3147-hcn}.
 The CO(1-0)/HCN(1-0) ratio is equal to 20  on average.
 Because the observed region contains the AGN, we would expect enhanced
 HCN emission there.
 Hence, the observed CO(1-0)/HCN(1-0) ratio is unexpectedly high.
 For instance in NGC 1097, this ratio is 3 in the nucleus
 and 10 in the star-forming ring \citep{kohno}. NGC~3147 is more similar 
 to the starburst in NGC~6951 \citep[][]{krips07b},
 where the ratio is 30 in the ring, and 2.5 in the nucleus.

 To study the physics and distribution of the molecular gas in detail 
 we need to analyze both single dish and interferometric observations.
 The analysis in the remainder of the paper will be performed on the 
 combined IRAM PdBI+30\,m datasets.

 \begin{figure*}
 \centering
 \includegraphics[width=0.43\textwidth,angle=-90,bb=145 17 460 750]{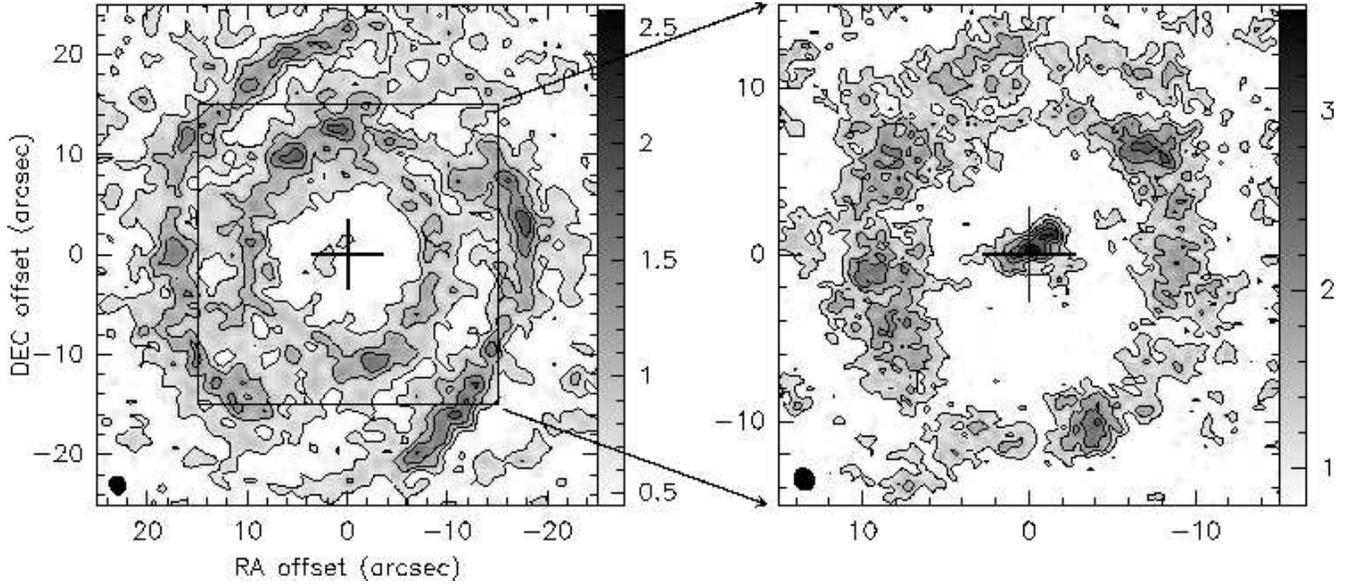}
 \caption{\textit{Left panel}: $^{12}$CO(1-0) integrated intensity 
 contours observed with the IRAM
 PdBI+30\,m toward the center of NGC\,3147.
 The cross marks the coordinates
 of the center as given in Table \ref{table1}, with offsets in arcseconds.
 The rms noise level is $\sigma = 0.15\,{\rm Jy\,beam^{-1}\,km\,s^{-1}}$.
 The map, derived with 2$\sigma$ clipping, has not been corrected for
 primary beam attenuation.  Contour levels run from 4$\sigma$ to 17$\sigma$
 with 2.5$\sigma$ spacing.  In this map the full $\pm 240\,{\rm km\,s^{-1}}$
 velocity range is used. The beam of 1\farcs9 $\times$ 1\farcs6
 is plotted at lower left.
 \textit{Right panel}: Same for $^{12}$CO(2-1).
 The rms noise level is $\sigma = 0.2\,{\rm Jy\,beam^{-1}\,km\,s^{-1}}$.
 Contour levels run from 5$\sigma$ to
 18$\sigma$ with 2.5$\sigma$ spacing.
 The beam of 1\farcs4 $\times$ 1\farcs2 is plotted at
 lower left.}
 \label{co10r}
 \end{figure*}

\begin{figure*}
 \centering
\hbox{
 \includegraphics[width=0.34\textwidth,angle=-90]{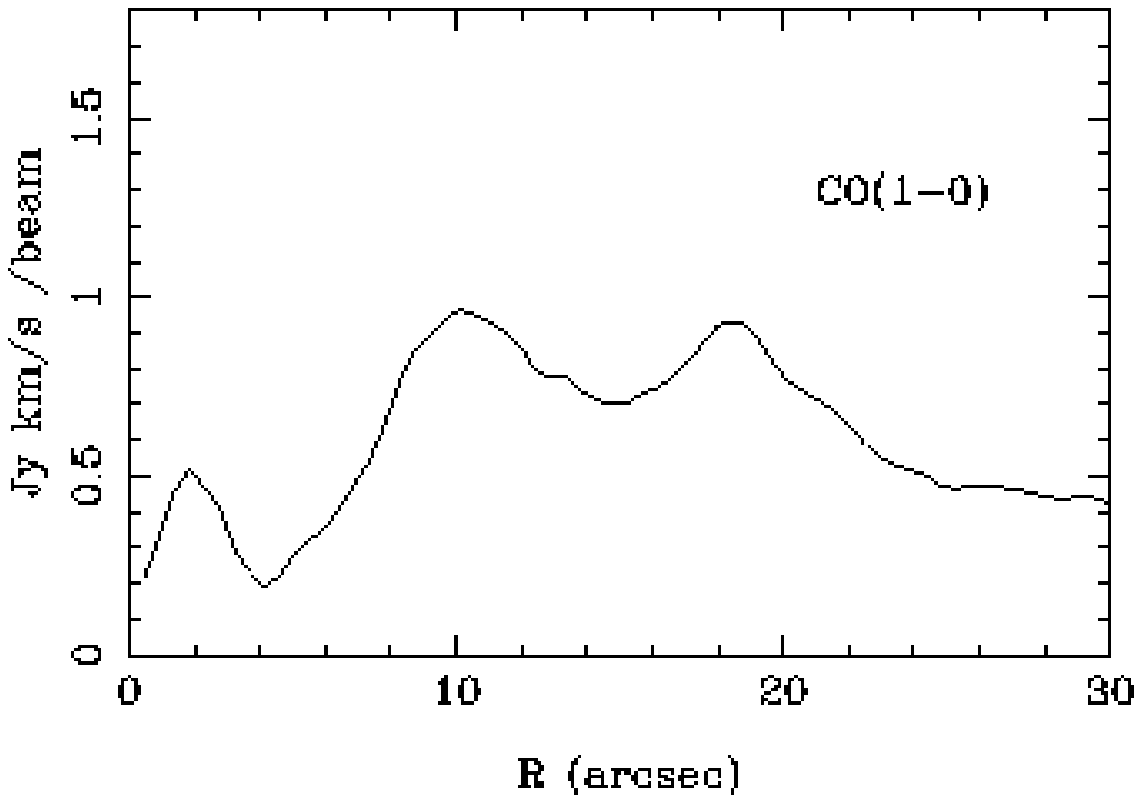}
\includegraphics[width=0.34\textwidth,angle=-90]{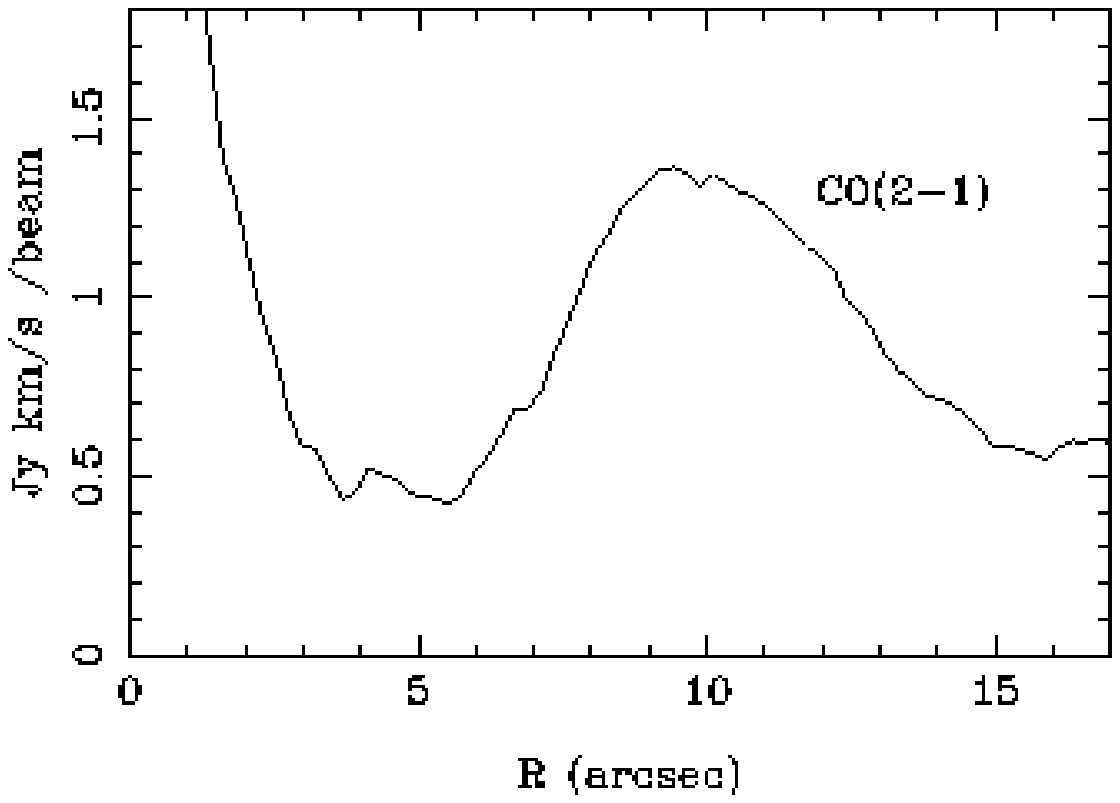}
}
 \caption{\textit{Left panel:} 
Radial distribution (azimuthal average, deprojected to face-on
 orientation) of the $^{12}$CO(1-0) integrated intensity shown in Fig. \ref{co10r}.
\textit{Right panel:} Same for  $^{12}$CO(2-1).
}
 \label{moment1021}
 \end{figure*}

 \section{Interferometric results: Molecular gas properties }

 \subsection{Morphology and mass of the CO rings\label{sec:morph}}

 The $^{12}$CO(1-0) and $^{12}$CO(2-1) integrated intensity distributions are  
 shown in Fig. \ref{co10r}.
The figure reveals a nuclear peak, a symmetric and 
 complete inner ring at a distance of about 10$^{\prime\prime}$ 
 from the nucleus, and a larger and incomplete ring at 
 a distance of about 20$^{\prime\prime}$ from the nucleus (left panel). 
 In $^{12}$CO(2-1) the nuclear peak is stronger and more visible (right panel),
 but there is no outer ring at a radius of 20$^{\prime\prime}$ 
 because of the restricted field-of-view (FWHM of primary beam 
 of 22$^{\prime\prime}$).
 The outer ring-like structure can also be interpreted as spiral arms, 
 more tightly wound in the inner regions than in the outer ones.
 The radial distribution of the azimuthally averaged $^{12}$CO(1-0) 
 and $^{12}$CO(2-1) intensities
 is shown in Fig. \ref{moment1021}, where there are clear dips 
 between the central peak and the inner ring and between the inner
 ring and the outer spiral structure (partial outer ring).

 Both $^{12}$CO(1-0) structures are clumpy, composed of individual giant 
 molecular cloud complexes,
 each with a mass of a few 10$^{7}$ to 10$^{8}$ M$_{\odot}$. The total
 $H_{2}$ mass derived from the interferometer maps, 
 assuming the same CO-to-H$_2$ conversion value used before, is 
 M$_{H_{2}}$ = 3.8 $\times$ 10$^9$ M$_{\odot}$. 
 The inner ring alone contributes more than half 
 of the total masses; M$_{H_{2}}$ = 2 $\times$ 10$^9$ M$_{\odot}$,
 and the two parts of the outer spiral have similar mass, 
 the western part has a mass of M$_{H_{2}}$ = 9.5 $\times$ 10$^8$ 
 M$_{\odot}$ and the eastern one
 M$_{H_{2}}$ = 8.5 $\times$ 10$^8$ M$_{\odot}$.
 The total M$_{H_{2}}$  (= 3.8 $\times$ 10$^9$ M$_{\odot}$) 
 we find is in good agreement with the mass 
 derived by \citet{young} for a position centered on 
 the galaxy nucleus, M$_{H_{2}}$ = 4.4 $\times$ 10$^9$ M$_{\odot}$.
 Our mass estimate from single dish observations 
 presented in Sect. \ref{singler} is larger, since it corresponds
 to a more extended region. 
 In any case, this molecular gas mass is very large, more than
 three times larger than the most massive of
 the other NUGA galaxies studied so far 
 \citep[NGC\,2782 and NGC\,4569, see][]{hunt,boone07}.

\begin{figure}
 \centering
\includegraphics[width=0.85\columnwidth,angle=-90]{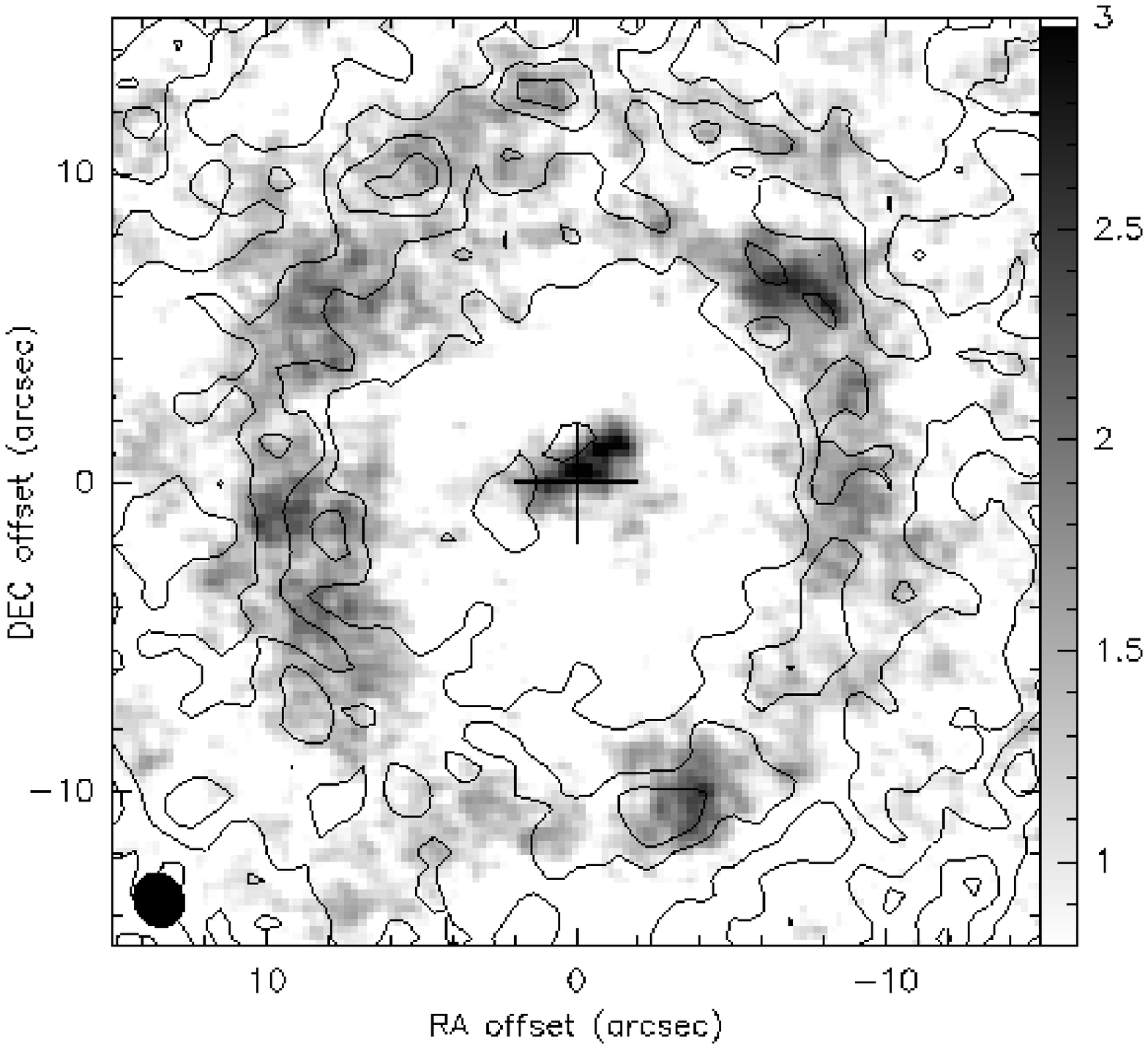}
 \caption{$^{12}$CO(1-0) contours as in Fig. \ref{co10r} (left panel) superposed on the 
 greyscale $^{12}$CO(2-1) map that has been tapered, convolved to the same resolution, 
 and corrected for primary beam attenuation, in units of ${\rm Jy\,beam^{-1}\,km\,s}^{-1}$. 
 The beam is plotted at lower left.}
 \label{co10-21}
 \end{figure}

\begin{figure}
	\centering
\includegraphics[width=0.85\columnwidth,angle=-90]{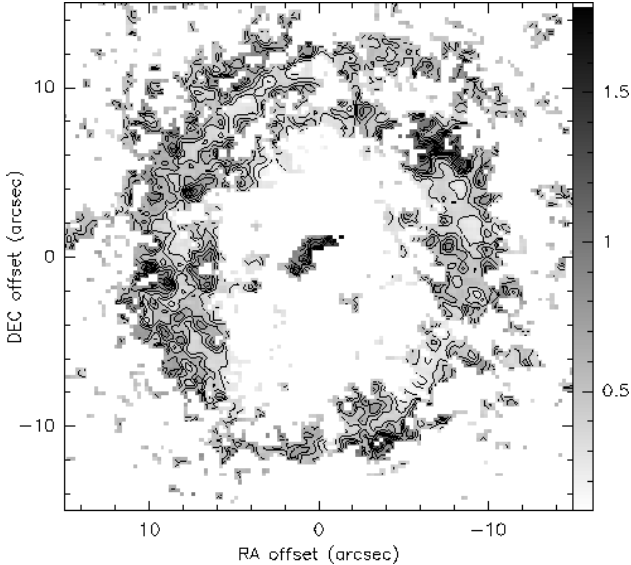}
	\caption{Contours and greyscale of the CO(2-1)/CO(1-0) ratio map.
Contours run from 0.1 to 1.8 in steps of 0.1 in temperature units.}
	\label{ratio}
  \end{figure}

 The comparison between the $^{12}$CO maps of the two transitions, 
 at the same resolution and with the same spatial frequency sampling, 
 gives insight about the excitation conditions 
 of the molecular gas locally, pixel by pixel.
 Fig. \ref{co10-21} shows that when the $^{12}$CO(2-1) data are tapered
 and convolved to the $^{12}$CO(1-0) resolution, the maxima of the $^{12}$CO(2-1) ring agree 
 quite well with the $^{12}$CO(1-0) peaks.
 The ratio map is shown in Fig. \ref{ratio}
 where the values range from 0.1 to 1.8.
 The mean ratio is $\sim$0.7, consistent with the value found 
 with single dish observations
 and also, in this case, consistent with the bulk of emission 
 being optically thick, as expected
 in the nuclei of spiral galaxies.
 Some regions, especially the center, reach a higher ratio ($\sim$1.3-1.4)
 where the gas  is warm and dense, while in the disk and spiral arms
 it is colder and more diffuse.
 The peak in the northwest region of the CO(2-1)/CO(1-0) ratio map
 is coincident  with the strongest HCN(1-0) emission corresponding 
 to the offset (-7$^{\prime\prime}$,7$^{\prime\prime}$).

 \subsection{Kinematics \label{sec:kinematics}}

\begin{figure}[ht]
	\centering
       \includegraphics[width=0.85\columnwidth,angle=-90]{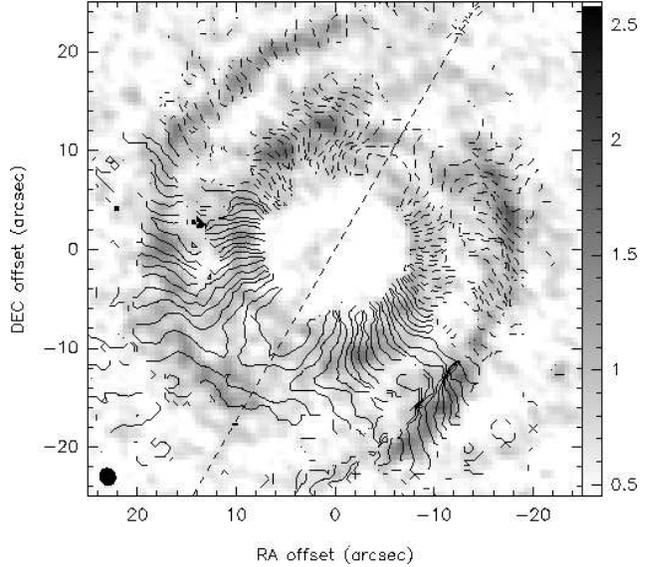}
	\caption{Overlay of the integrated $^{12}$CO(1-0) emission in grey scale,
	same as Fig. \ref{co10r} (left panel), with the CO mean-velocity field in contours 
	spanning the range -200 km s$^{-1}$ to 200 km s$^{-1}$ in steps of 
	10 km s$^{-1}$. The velocities are referred
	to V$_{hel}$ = 2813 km s$^{-1}$. Solid (dashed) lines are used for positive
	(negative) velocities. The dotted line indicates the major axis (PA = 150$^\circ$).}
	\label{co-velo}
 \end{figure}

 The channel maps in Fig. \ref{channel} display overall a gross 
 regularity of the large scale kinematics, 
 following the expected spider diagram.
 However, there are some local wiggles superposed on this pattern, 
 i.e., streaming motions.
 Such streaming motions are also visible in 
 Fig. \ref{co-velo}, which shows the isovelocity curves (first-moment map) of the 
 $^{12}$CO(1-0) emission superposed on the $^{12}$CO(1-0) integrated intensity. 
 Some of the perturbations in the velocity field are 
 clearly coincident with the spiral arms, as expected from density wave theory.

\begin{figure}[h]
	\centering
\includegraphics[width=0.95\columnwidth]{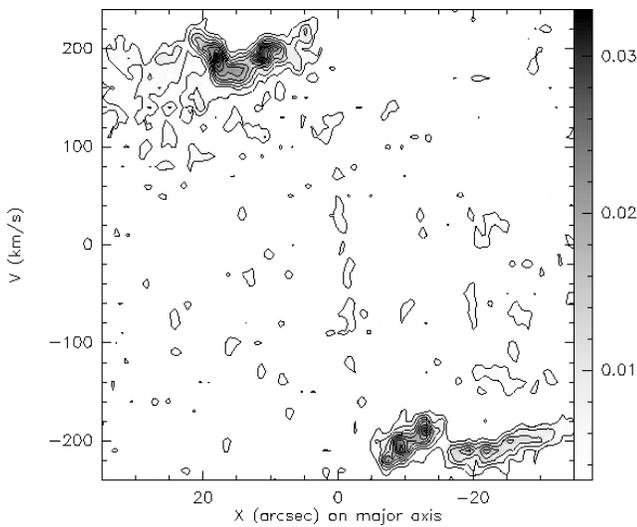}
	\caption{$^{12}$CO(1-0) position-velocity (p-v) diagram along the major
	axis of NGC~3147 using the whole velocity range +/-240 km s$^{-1}$.
	The coordinates of the center (0,0), the heliocentric velocity, and PA used are
	given in Table \ref{table1}. Contour levels are from 3.2 to
  $32\,{\rm mJy\,beam^{-1}}$ in steps of $3.2\,{\rm mJy\,beam^{-1}}$.}
	\label{p-v}
\end{figure}

 A position-velocity (p-v) cut along the major axis using a position angle of
 150$^\circ$ is
 presented in Fig. \ref{p-v}.
 The major-axis p-v diagram shows approximately regular kinematics,
 although there are clear signs of deviations from ``flatness'' at $\sim$15\arcsec
 on either side of the nucleus of amplitude $\sim$40\,km\,s$^{-1}$.

\begin{figure}[h]
	\centering
\includegraphics[width=0.95\columnwidth]{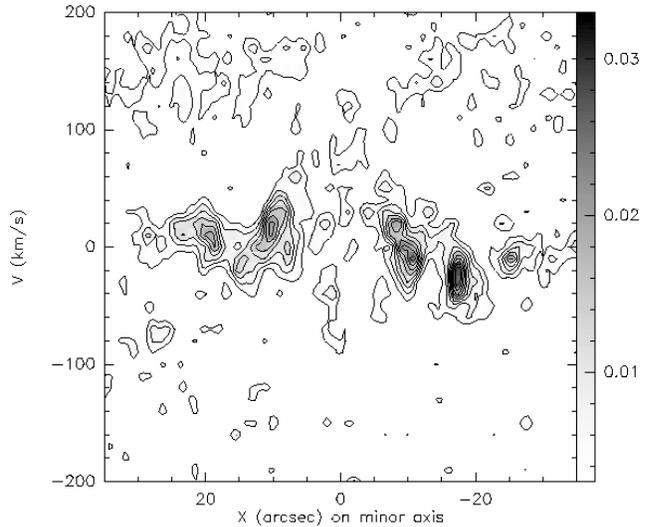}
	\caption{$^{12}$CO(1-0) position-velocity (p-v) diagram along the minor
	axis of NGC~3147 using the same velocity range, center, PA, and 
        systemic velocity as for Fig. \ref{p-v}.
        Contour levels are from 2.8 to $28\,{\rm mJy\,beam^{-1}}$ in steps of 
        $2.8\,{\rm mJy\,beam^{-1}}$.}
	\label{fig:minor}
        \end{figure}

  The minor-axis p-v diagram is displayed in Fig. \ref{fig:minor}.
  The wiggles are more conspicuous on the major axis in Fig. \ref{p-v}
  than in Fig. \ref{fig:minor}.
  Although the minor-axis velocities are 
  not completely symmetric about the nucleus, they hint that 
  the outer spiral arm/pseudo-ring at a radius of 20\arcsec\ 
  is already outside corotation.
  This is because the sign of the radial streaming motions 
  (visible on the minor-axis) is expected to change at corotation, while 
  the tangential streaming (visible on the major-axis) does not change 
  at corotation \citep[e.g.][]{canzian93}. 
  Thus, we conclude (although with substantial
  uncertainty) that corotation occurs at $\sim$15\arcsec\ (3\,kpc), 
  consistent with the minor-axis p-v plot. This result will help constrain
  the pattern speed of the weak bar/oval which we discuss in
  Sect. \ref{sec:discussion}.

 \subsection{Dynamical mass \label{sec:mass}}

 The peak velocity of $\sim$200\,km\,s$^{-1}$
 in the major-axis p-v diagram occurs
 at a radius of $\sim$7-8\,\arcsec\ (1.5\,kpc).
 This value, when corrected for disk
 inclination (we adopt $i=29.5^\circ$, see Table \ref{table1}) by a 
 factor sin~$i$,
 becomes 400\,km\,s$^{-1}$.
 From the corrected peak velocity, we can estimate the
 dynamical mass within this radius:
 $M(R) = \alpha \times {\frac {R V^2(R)}{G}}$,
 where $M(R)$ is in $M_\odot$, $R$ in kpc, and $V$ in km\,s$^{-1}$.
 Assuming the most flattened disk-like distribution
 ($\alpha = 0.6$),
 gives a dynamical mass $M_{\rm dyn}=3.3\times10^{10}$\,$M_\odot$
 within a diameter of $\sim$3\,kpc.
 Assuming a roughly flat rotation curve (see Sect. \ref{sec:discussion}),
 this would imply a mass $M_{\rm dyn}=8.9\times10^{10}$\,$M_\odot$
 within 4\,kpc (roughly the radius of the PdBI primary
 beam for our $^{12}$CO(1-0) observations).
 With a total molecular gas mass of $\sim4\times10^{9}$\,$M_\odot$
 (see Sect. \ref{sec:morph}),
 we would thus estimate a molecular mass fraction of $4-5$\%.

\section{Comparison with other wavelengths \label{sec:other}}

 It is instructive to compare the distribution of the 
 molecular gas in the inner kpc of NGC~3147 with observations at 
 other wavelengths. 
 In particular, molecular clouds  are thought to be the birth site of
 future generations of stars, and thus should be strongly connected to other 
emission  tracing  
 star formation, such as far-ultraviolet and warm dust emission.

\subsection{The bar and the lopsided nucleus \label{sec:bar}}

\begin{figure*}
\includegraphics[width=0.43\textwidth,angle=-90]{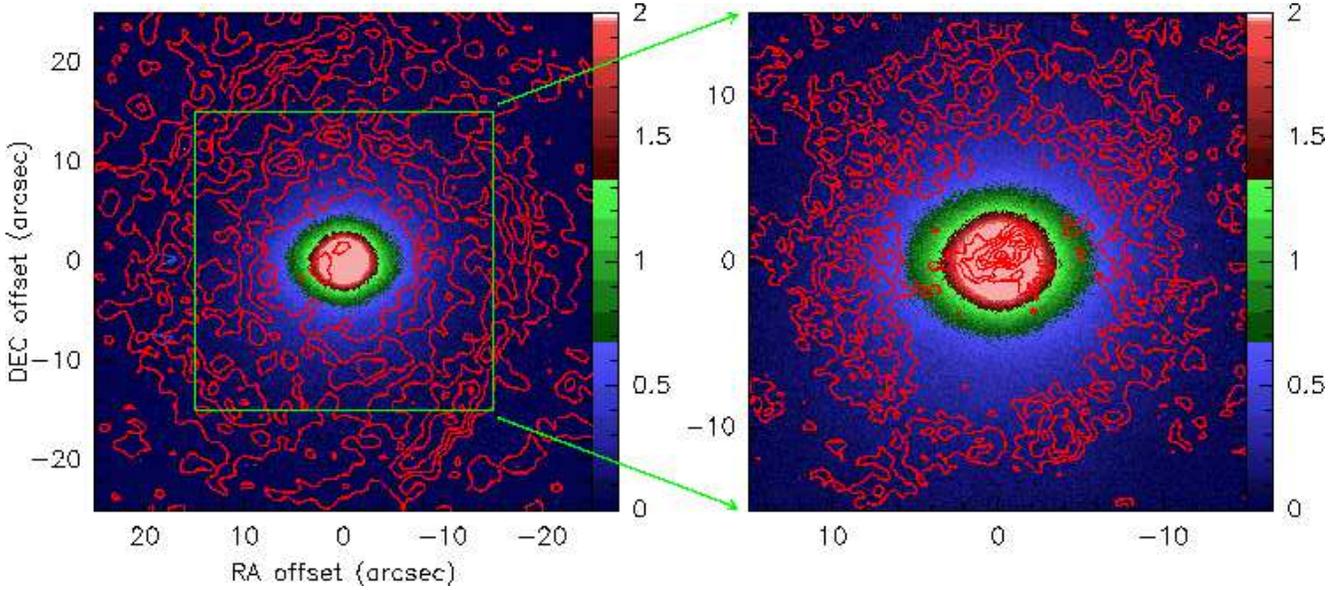}
	\caption{\textit{Left panel}: $^{12}$CO(1-0) contours 
         (0.6 to 2.6 by 0.35 Jy km s$^{-1}$ beam$^{-1}$)
         superposed on the near-infrared $J$ image from the CFHT,
         shown in logarithmic levels. The central 50$^{\prime\prime}$ are
	 shown. \textit{Right panel}: Same for $^{12}$CO(2-1) contours 
         (1.0 to 3.5 by 0.5 Jy km s$^{-1}$ beam$^{-1}$).
        The central 30$^{\prime\prime}$ are
	shown.
        }
        \label{JCO}
        \end{figure*}

The CO contours are superposed on the CFHT $J$-band image in Fig. \ref{JCO}.  
There is a weak bar/oval feature at a position
angle PA = 85$^\circ$,  just contained inside the inner CO
ring. Only the NIR reveals this bar, since
NGC~3147 is not classified optically as a barred galaxy.  Its 
late type (Sbc) corresponds to a non-dominating bulge,
and the bar feature is likely not diluted by this bulge.
The major axis orientation differs by 65$^\circ$ from the bar
PA, which minimizes confusion due to projection effects.

\citet{mh2001}
  have obtained $JHK$ images of NGC\,3147 with the 2.2\,m telescope 
  of the Calar 
  Alto observatory using the MAGIC NICMOS3 camera, together with 40 other 
  unbarred galaxies with low inclinations. They
  decomposed the images into bulge and disk components, and noticed that 
  the bulge in NGC~3147 appears at a very different position angle than the disk;
  they concluded that the bulge must be triaxial, which was a rare 
  feature in their sample. 
 We have re-reduced their MAGIC $J$ and $K'$ images and the oval is present,
 similar to the feature seen in our CFHT images, although with noisier 
 isophotes.
 Our kinematic data suggest that
 this bar may be the agent of the inflow that we discuss in
 Sect. \ref{torq},

 The ellipticity profile shown in Fig. \ref{fig:ellipse} (see Sect. \ref{sec:archive})
 helps us estimate the length of the bar/oval distortion.
 The ellipticity peaks at $\sim$5\arcsec\ (1\,kpc), but the bar
 extends roughly to a radius of $\sim$7.5\arcsec\ (1.5\,kpc).
 This is also seen in the $J$-band image (Fig. \ref{JCO}) where the bar is roughly
 16\arcsec\ in diameter, and in the Fourier decomposition of the 
 stellar potential discussed in Sect. \ref{torq}.
 We therefore adopt a bar radius of $\sim$7.5\arcsec\ or $\sim$1.5\,kpc, 
 although this estimate is subject to a rather large uncertainty, $\pm$30\%.

\begin{figure}
 \centering
 \includegraphics[width=0.87\columnwidth,angle=-90]{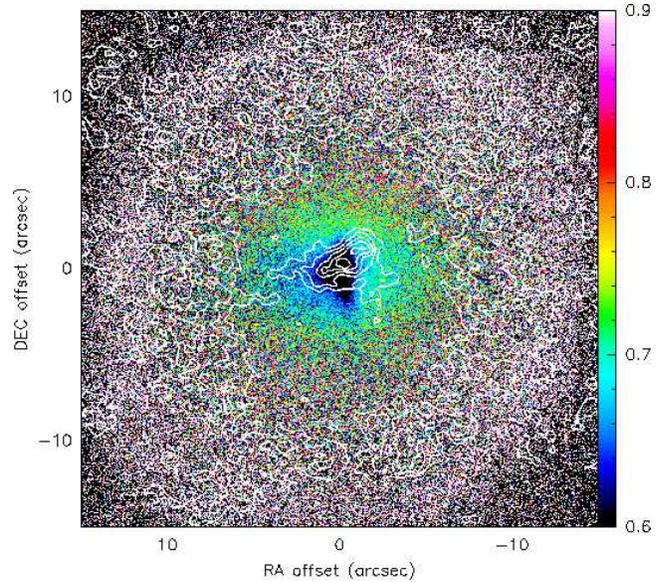}
 \caption{The $J-K$ image from the CFHT with
 $^{12}$CO(2-1) integrated intensity overlaid in contours
 (1.0 to 3.5 by 0.5 Jy km s$^{-1}$ beam$^{-1}$).
 North is up and east to the left.
 }
 \label{fig:jk}
 \end{figure}

 The $J-K$ image of NGC~3147
 is shown in Fig. \ref{fig:jk} with the $^{12}$CO(2-1) integrated
 intensity overlaid as contours (see Sect. \ref{sec:morph}).
 Coincident with the nuclear CO emission,
 there is a clearly-defined
 non-axisymmetric structure of $\sim$700\,pc in size, which is bluer 
 in $J-K$ than the surrounding regions that are devoid of gas. 
 This central asymmetric ``blob'' or ``lopsided nucleus''
 could be the destination of the gas inflow that we discuss in Sect. \ref{torq}.

 \subsection{The inner ring and the outer spiral \label{sec:ring}}

\begin{figure}%[!t]
 \centering
 \includegraphics[width=0.9\columnwidth,angle=-90]{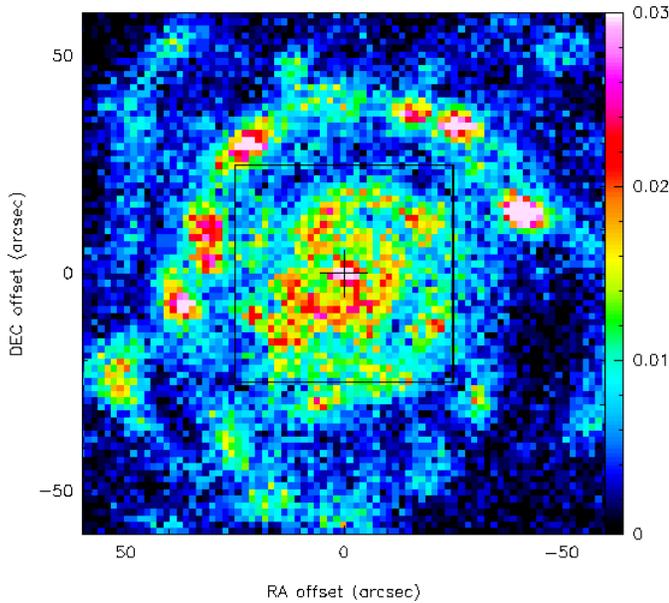}
 \caption{\textit{GALEX} FUV sky subtracted image of NGC~3147.
 The central 120$^{\prime\prime}$ of the galaxy are
 shown.
 The box shows the central 50$^{\prime\prime}$, 
 field of view of our $^{12}$CO observations.
 }
 \label{onlygalex}
 \end{figure}

 Fig. \ref{onlygalex} displays the
 FUV image obtained with the \textit{GALEX} satellite, described in
 Sect. \ref{sec:archive}.
 Inspection of Fig. \ref{onlygalex}
 shows that NGC~3147 has a quite well-defined outer spiral 
 structure, outside the range of our CO observations.
 There is also a compact nucleus with a central peak,
 surrounded by a roughly axisymmetric ``plateau'' of emission
 which breaks up into a pseudo ring at a radius of $\sim$20\arcsec.

\begin{figure*}
 \centering
\includegraphics[width=0.43\textwidth,angle=-90]{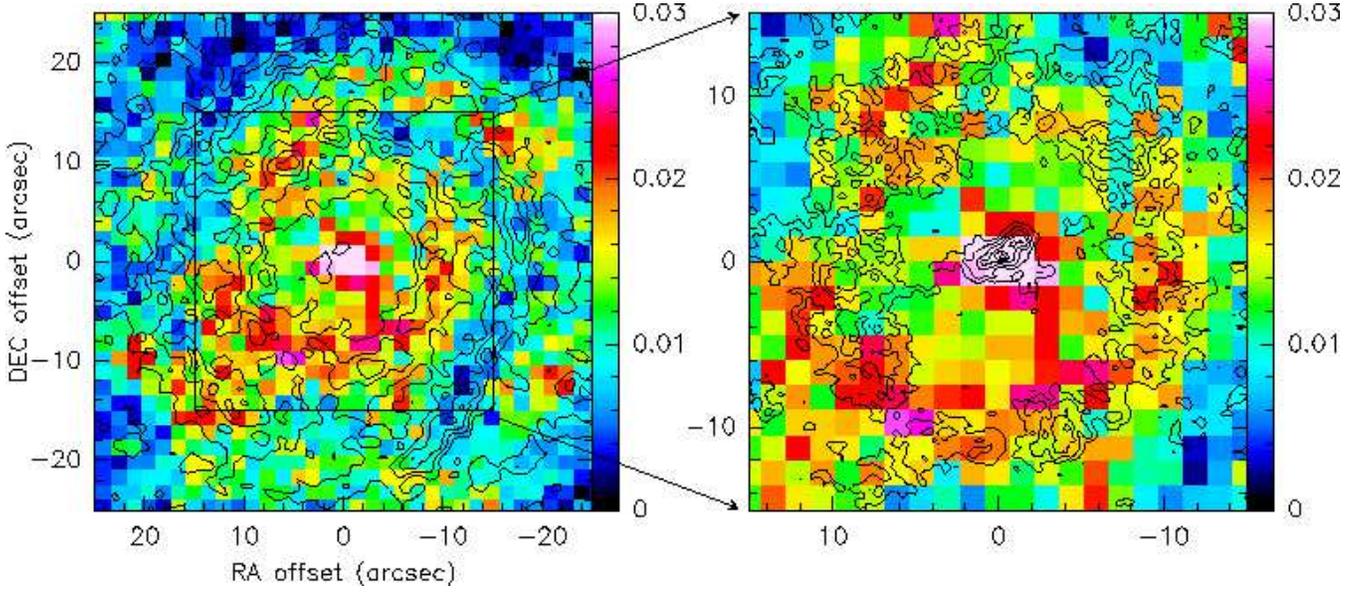}
 \caption{\textit{Left panel}: $^{12}$CO(1-0) contours 
         (0.6 to 2.6 by 0.35 Jy km s$^{-1}$ beam$^{-1}$) superposed on 
         the \textit{GALEX} FUV image of NGC~3147, shown in logarithmic 
         levels. The central 50$^{\prime\prime}$ are
         shown. \textit{Right panel}: Same for $^{12}$CO(2-1) contours  
         (1.0 to 3.5 by 0.5 Jy km s$^{-1}$ beam$^{-1}$). The central 
         30$^{\prime\prime}$ are shown.}
	\label{galex-co}
  \end{figure*}

 Fig. \ref{galex-co} shows the superposition of \textit{GALEX} FUV 
 image with the $^{12}$CO(1-0) (left panel) and $^{12}$CO(2-1) (right) 
 integrated intensity  contours.
 The cross shown in Fig. \ref{onlygalex} is at 
 the center of our CO observations, and coincides with the peak of the FUV nuclear emission. 
 This agreement suggests that the FUV and CO emission are cospatial 
 in the nuclear region.

 However, there is no clear ring in the FUV image.
 The inner 10$^{\prime\prime}$-radius 
 CO ring is contained inside the circular region of the FUV emission, and the outer 
 CO spiral (20$^{\prime\prime}$-radius), at least its eastern part,  has
 some overlap with the FUV spiral, but there are clear shifts between the two distributions. 
 The $^{12}$CO(2-1) emission (essentially the inner ring) also appears to settle 
 inside the \textit{GALEX} FUV inner ``plateau'' (Fig. \ref{galex-co}).

 The ``outer spiral'' in $^{12}$CO(1-0) falls in the interarm ``gap''
 in the \textit{GALEX} image.
 Offsets between FUV and CO (or FIR and HI) emissions are not 
 uncommon in grand design spiral galaxies, such as M\,100 
 \citep{rand95,sempere97} and M 51 \citep{calzetti05}, 
 and may relate to star-formation efficiency and timescale variations 
 in response to a spiral density wave.
 The star formation and all related tracers are often located  
 in different regions of the spiral arms: FUV emission is more prominent 
 at the outer edge of the arms, where the dust extinction is low, while  
 FIR and ${\rm H\alpha}$ emission is stronger at the inner edge.
 However, in NGC 3147 we find CO in the middle 
 of the interarm region in the \textit{GALEX} image, not at the inner or 
 outer edge of the arm; this result is quite rare.

 Unlike the FUV, the 8\,$\mu$m ``dust-only''
 image described in Sect. \ref{sec:archive} shows an inner ring
 apparently associated with molecular gas.

\begin{figure}
	 \centering
                 \includegraphics[width=0.9\columnwidth,angle=-90]{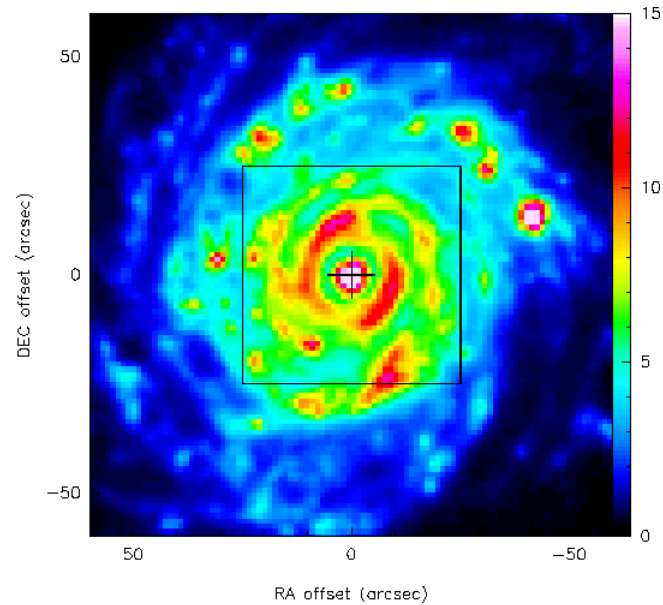}
\caption{\textit{Spitzer/IRAC} ``dust-only'' image at 8\,$\mu$m of NGC~3147.
	  The central 120$^{\prime\prime}$ of the galaxy are shown.
The box shows the central 50$^{\prime\prime}$, 
 field of view of our $^{12}$CO observations.
}
	 \label{spitzer-8um}
  \end{figure}

\begin{figure*}
	\centering
\includegraphics[width=0.43\textwidth,angle=-90]{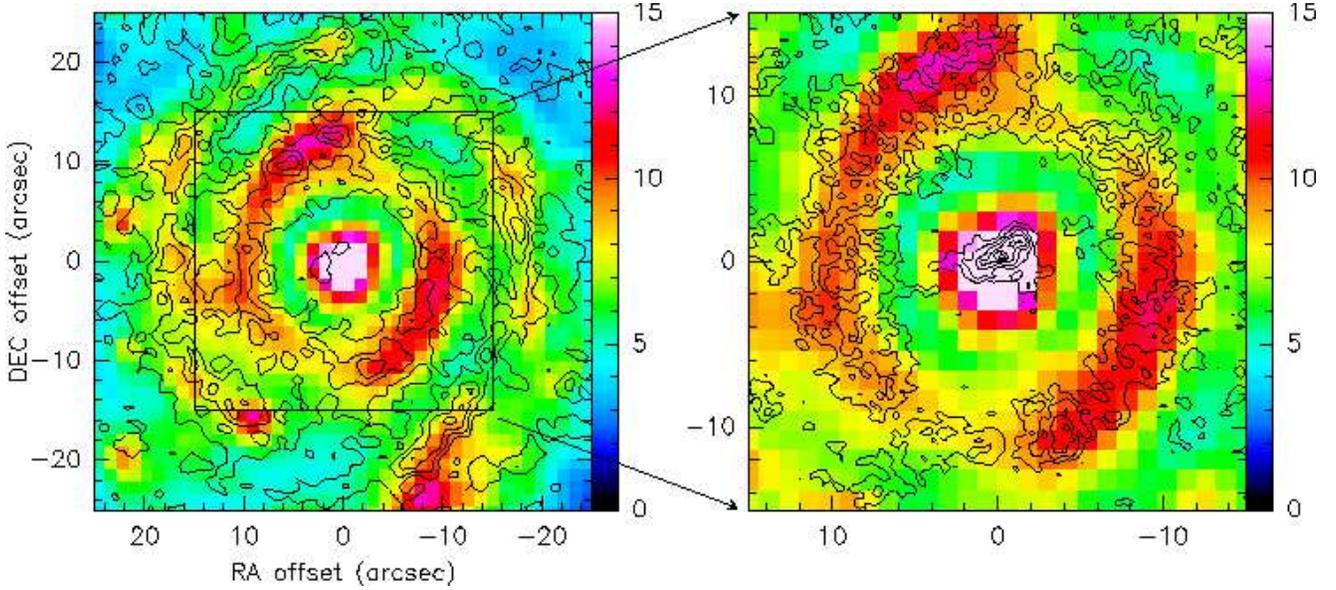}
   \caption{\textit{Left panel}: $^{12}$CO(1-0) contours 
    (0.6 to 2.6 by 0.35 Jy km s$^{-1}$ beam$^{-1}$) superposed on 
    the 8\,$\mu$m ``dust-only'' (\textit{Spitzer/IRAC}) image of 
    NGC~3147, shown in logarithmic levels. The central 50$^{\prime\prime}$ are
    shown. \textit{Right panel}: Same for $^{12}$CO(2-1) contours  (1.0 to 3.5 by 
    0.5 Jy km s$^{-1}$ beam$^{-1}$). The central 30$^{\prime\prime}$ are
    shown.}
   \label{spitzer-co}
   \end{figure*}

 Fig. \ref{spitzer-8um} shows this image,
 which clearly exhibits an inner-ring structure at about 10$^{\prime\prime}$ radius, 
 together with a central peak. 
 Such rings are frequently observed in barred galaxies with 
 \textit{Spitzer} \citep[e.g.][]{regan06}.
 Fig. \ref{spitzer-co} 
 shows the dust-only 8\,$\mu$m image with the $^{12}$CO(1-0) (left panel) and $^{12}$CO(2-1)
 (right) intensity contours superposed.
 The ring visible at 8 $\mu$m corresponds to both the $^{12}$CO(1-0) 
 inner ring and the $^{12}$CO(2-1) one. 
 However, the outer $^{12}$CO(1-0) spiral is slightly inside the corresponding
 structure in dust emission. 
 This may be a similar phenomenon to that seen in the \textit{GALEX} FUV
 image, where molecular gas peaks in the interarm regions as described
 above.

 We can use the 8\,$\mu$m dust-only image, together with
 the CO emission itself, to define the size and width of the inner
 gas (and dust) ring.
 The ring is rather broad, extending over roughly 2--3\arcsec\ in radius,
 or $\sim$500\,pc.
 Measured to the middle of its width, the inner gas ring has a radius
 of $\sim9-10$\arcsec\ (1.8--2\,kpc) , although it is slightly elongated, being
 longer along the horizontal axis.
 If instead we measure to the innermost portion of the ring, we
 find a radius of $\sim7^{\prime\prime}-8^{\prime\prime}$ ($\sim$1.5\,kpc).
 This uncertainty (20-30\%) will influence our discussion in Sect. \ref{sec:discussion},
 where we will use the size of the ring in the context of the rotation
 curve to help constrain the pattern speed and location of the resonances
 that define the kinematics in NGC~3147.

\subsection{Star formation \label{sec:sf}}

 The agreement, where present, between the molecular gas distribution and the 
 ultraviolet and infrared emission is consistent with on-going star formation 
 within the molecular clouds. 
 The star formation rate (SFR) can be computed from both 
 UV and FIR emission using the calibrations given by \citet{kennicutt98} 
for a $0.1-100\,M_\odot$ Salpeter initial mass function:

 \begin{eqnarray}
 \mbox{SFR} \,\mbox{[M$_{\odot}$\,yr$^{-1}$]} = 1.4 \times 10^{-28}\,\mbox{L$_{UV}$}	
 \label{sfr1}
 \end{eqnarray}
 where L$_{UV}$ is the UV luminosity in
 ergs s$^{-1}$ Hz$^{-1}$ over the wavelength range 1500-2800 $\AA$, and 

 \begin{eqnarray}
 \mbox{SFR} \,\mbox{[M$_{\odot}$\,yr$^{-1}$]} = 4.5 \times 10^{-44}\, \mbox{L$_{FIR}$}	
 \label{sfr2}
 \end{eqnarray} 
 where L$_{FIR}$, expressed in ergs s$^{-1}$, refers to the total infrared luminosity 
 integrated over the entire mid and far-IR spectrum
 (8-1000 $\mu$m). 
 We first compute the UV luminosity from the \textit{GALEX} 
 map inside the FOV of 42$^{\prime\prime}$, to better separate the SFR in 
 the PdBI FOV from the total, and find a SFR $\sim$1\,M$_{\odot}$\,yr$^{-1}$.

 We then use the \textit{Spitzer} MIPS images available
 from the archive to estimate
 how much FIR flux falls within the FOV of the PdB primary beam
 compared to the total integrated over the much larger \textit{IRAS} beams.
 Rough measurements give a fraction of $\sim$12\% for 160\,$\mu$m and
 $\sim$22\% for 24\,$\mu$m (the 70\,$\mu$m image is not available).
 We then estimate the FIR luminosity by scaling the IRAS 100\,$\mu$m and 60\,$\mu$m fluxes
 by 12\% and 17\%\footnote{17\% is a rough interpolation of the central concentration
 we would expect at 60\,$\mu$m, given the more compact emission at 24\,$\mu$m
 and the more extended emission at 160$\mu$m.}, respectively.
 This gives 1.39\,Jy at 60\,$\mu$m and 3.55\,Jy at 100\,$\mu$m, which results in
 an estimate for the SFR of $\sim$0.8\,M$_{\odot}$\,yr$^{-1}$,
 similar to that obtained from \textit{GALEX}.

 The compatibility between the two estimations of SFR (from UV and FIR) means that there is not
 significant obscuration, which is probably due to the nearly face-on view of the galaxy.

\begin{figure}
  \includegraphics[angle=-90,width=\columnwidth]{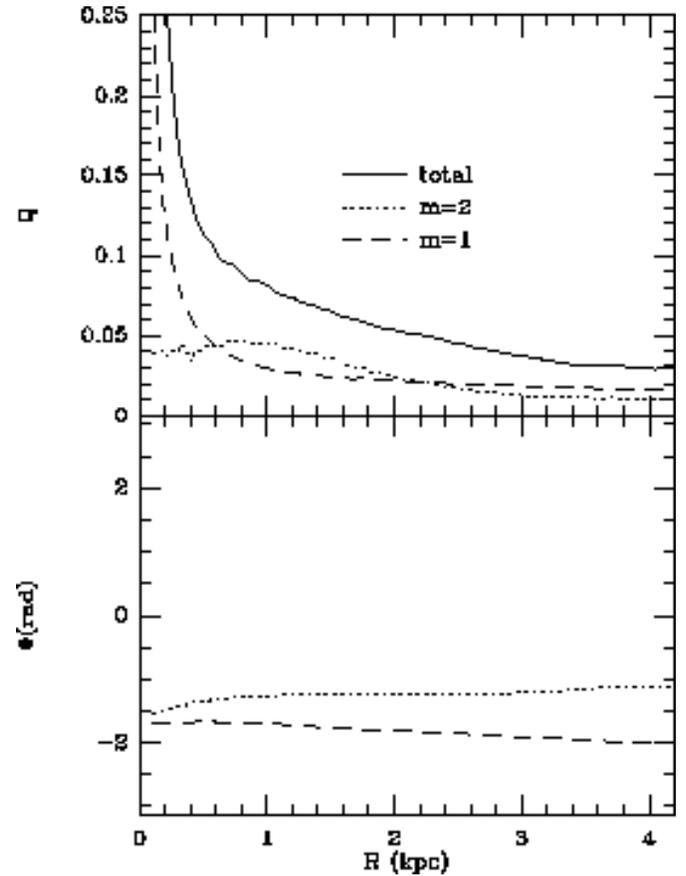}
  \caption{Strengths (Q$_1$, Q$_2$, and total Q$_T$) and phases ($\phi_1$ and $\phi_2$) of 
  the $m=1$ and $m=2$ Fourier
  components of the stellar potential inside a field of 
  48$^{\prime\prime}$ in diameter (r $<$ 4.6 kpc).}
  \label{pot3147}
  \end{figure}

 The SFR per unit area inferred from these values is 
 0.016\,M$_{\odot}$\,yr$^{-1}$\,kpc$^{-2}$,
 high for spiral disks but similar to, although slightly lower than, typical
 values for IR-selected circumnuclear starbursts \citep{kennicutt98}.
 Moreover, given the gas mass we derived in Sect. \ref{sec:morph}
 and the correlations shown in \citet{kennicutt98},
 the global star-formation efficiency within the inner 8\,kpc is low, $\sim$1\% rather
 than the more typical value of $\sim$10\%.
 On the other hand, the molecular gas is concentrated into well-defined
 structures; more than half the molecular mass is found in the inner ring
 alone (see Sect. \ref{sec:morph}).
 If we consider only the molecular gas in the inner ring, the gas surface
 density is $\sim$320\,M$_{\odot}$\,pc$^{-2}$, roughly 4 times higher
 than we obtain by averaging over the entire FOV.
 Nevertheless, it is significantly lower than that in M\,51 
 spiral arms \citep{henry03}.

\section{Computation of the torques \label{torq}}

  \subsection{Evaluation of the gravitational potential \label{grav-pot}}

  As described in previous papers \citep[e.g.,][]{garcia05},
  we assume that NIR images can give us the best approximation
  for the total stellar mass distribution,  being less affected than 
  optical images by 
  dust extinction or by stellar population biases.  We here recall
  the essential definitions and assumptions. 

  The $J$-band image was first deprojected according to the
  angles PA = 150$^\circ$ and $i=29.5^\circ$. We have not deprojected
  the bulge separately, since we do not know its actual flattening, but 
  because the galaxy is of late type (Sbc), it does not have 
  a large contribution. The image is 
  completed in the vertical
  dimension by assuming an isothermal plane model with a constant scale height,  
  equal to $\sim$1/12th of the radial scale-length of the image. The potential is
  then derived by a Fourier transform method, assuming a constant mass-to-light (M/L) ratio. 
  The M/L value is selected to reproduce the observed CO rotation curve.
  The potential was also calculated from the lower-resolution
  MAGIC images, which have a much larger field-of-view (162\arcsec).
  The same results are obtained, and the potential is shown to be axisymmetric 
  at large radii.
  Beyond a radius of 20\arcsec, 
  the mass density
  is set to 0, thus suppressing any spurious $m=4$ terms; 
  this is sufficient to compute the potential over the PdB $^{12}$CO(1-0) primary beam.

For the non-axisymmetric part, the
potential $\Phi(R,\theta)$ is then decomposed in the different $\textit{m}$-modes:
  $$
  \Phi(R,\theta) = \Phi_0(R) + \sum_{m=1}^\infty \Phi_m(R) \cos (m \theta - \phi_m(R))
  $$
  \noindent
  where $\Phi_m(R)$ and $\phi_m(R)$ represent the amplitude and phase of the $m$-mode.
  
  The strength of the $m$-Fourier component, $Q_m(R)$ is defined as
  $Q_m(R)=m \Phi_m / R | F_0(R) |$, i.e. by the ratio between tangential
  and radial forces \citep[e.g.][]{combes81}.
  The strength of the total non-axisymmetric perturbation is defined by:
  $$
  Q_T(R) = {F_T^{max}(R) \over F_0(R)} 
  $$
  \noindent
  where $F_T^{max}(R)$ represents the maximum amplitude of the tangential force and $F_{0}(R)$ is the mean 
  axisymmetric radial force.
  Fig.~\ref{pot3147} displays these $Q$ values as a function of radius for NGC~3147.
  A bar ($m=2$) can be seen clearly, with a peak radius of $\sim$ 1\,kpc (5\arcsec),  
  and a total extent of 1.5\,kpc (7.5\arcsec) in radius.
  These dimensions are similar to those inferred from the
  ellipticity distribution discussed in Sect. \ref{sec:bar}, and strengthen
  our conclusion that this structure is ``bar-like''.
  There is also an asymmetry in the $m=2$ profile towards
  the center. The lopsidedness has a correspondence in the gas
  morphology described in Sect. \ref{sec:morph} and the $J-K$ image
  discussed in Sect. \ref{sec:archive}. 
  The bar is regular in phase, but its strength is relatively modest.

 \begin{figure}
  \includegraphics[angle=-90,width=\columnwidth,bb=80 80 530 610]{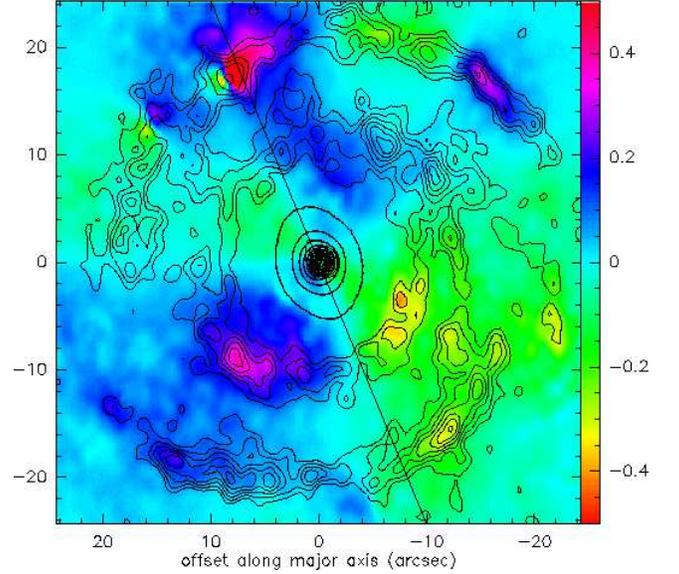}
\caption{CO(1--0) contours overlaid on the gravitational torque
  map (t(x,y)~$\times$~$\Sigma$(x,y), as defined in text) in the center of NGC~3147.
  The torque map (grey/color scale) is
  normalized to the maximum absolute value of the torques.
  The derived torques change sign as expected in a {\it butterfly} diagram,
  delineating four quadrants. The orientation of quadrants follow
  the bar orientation in NGC~3147. In this deprojected picture,
  the major axis of the galaxy is oriented parallel to the abscissa.
  The line reproduces the mean orientation of the bar
  (PA = 20$^\circ$ on the 
  deprojected image and 85$^\circ$ on the projected one).
  }
  \label{torq3147}
  \end{figure}

  \subsection{Evaluation of gravity torques}
\begin{figure}
  \includegraphics[angle=-90,width=\columnwidth,bb=80 80 530 610]{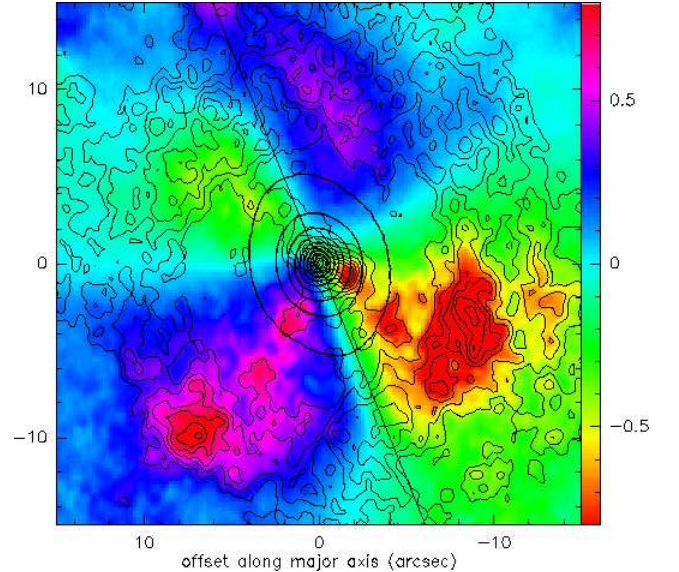}
  \caption{Same as Fig. \ref{torq3147} for the $^{12}$CO(2--1) 
  emission taken as a tracer of gas surface density.}
  \label{torq3147_2}
  \end{figure}

  After having calculated the forces per unit mass ($F_x$ and $F_y$) from the derivatives of $\Phi(R,\theta)$ on 
  each pixel, the torques per unit mass $t(x,y)$ can be computed by:
  $$
  t(x,y) = x~F_y -y~F_x
  $$
  The torque map is oriented according to the sense of rotation in the plane of the galaxy.
  The combination of the torque map and the gas density $\Sigma$  map then 
  allows us to derive the net effect on the gas at each radius.
  The gravitational torque map weighted by the gas surface density $t(x,y)\times
  \Sigma(x,y)$, normalized to its maximum value, is shown in Figs.~\ref{torq3147}
  and \ref{torq3147_2} for CO(1--0) and CO(2--1), respectively.

  To estimate the radial gas flow induced by the torques, we first compute the torque 
  per unit mass averaged over azimuth, using $\Sigma(x,y)$ as the actual weighting function, i.e.:
  $$
  t(R) = \frac{\int_\theta \Sigma(x,y)\times(x~F_y -y~F_x)}{\int_\theta \Sigma(x,y)}
  $$   

\begin{figure*}
  \includegraphics[angle=-90,width=\textwidth]{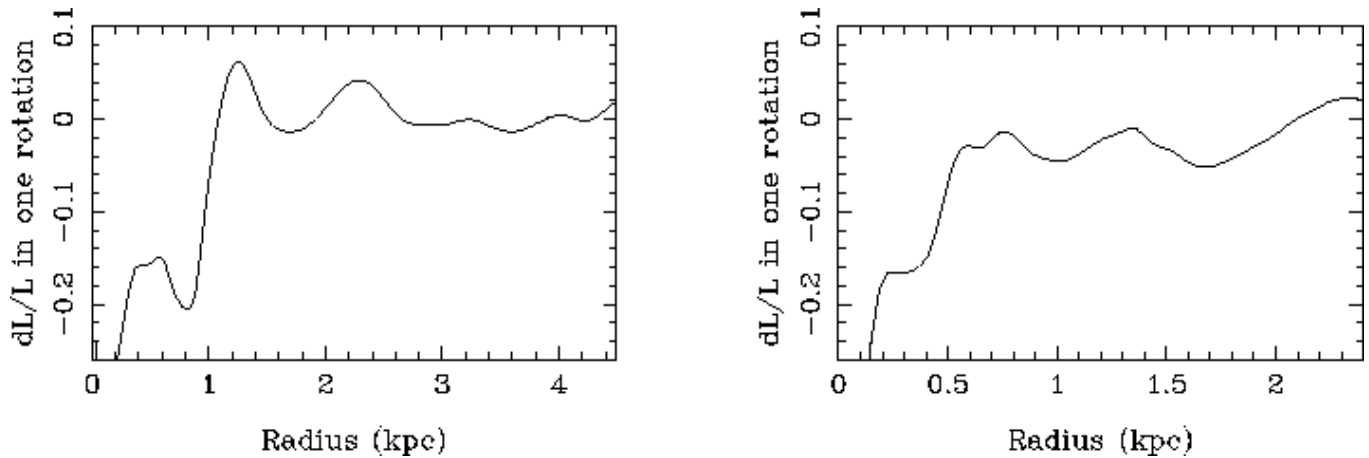}  
  \caption{The torque, or more precisely the fraction of the angular momentum transferred
  from/to the gas in one rotation --$dL/L$-- is plotted for $^{12}$CO(1-0) (left) and
  $^{12}$CO(2-1) (right).}
  \label{gastor3147}
  \end{figure*}

  By definition, $t(R)$ represents the time derivative of the specific angular momentum $L$ of the gas averaged 
  azimuthally, i.e., $t(R)$=$dL/dt~\vert_\theta$. To derive dimensionless  quantities,
  we normalize this variation of angular momentum per unit time
  to the angular momentum at this radius and to the rotation period.
  We then estimate the efficiency of the gas flow as
  the average fraction of the gas specific angular momentum transferred in one rotation 
  ($T_{rot}$) by the stellar potential, as a function of radius, i.e., by the function $\Delta L/L$ defined as:
  $$
  {\Delta L\over L}=\left.{dL\over dt}~\right\vert_\theta\times \left.{1\over L}~\right\vert_\theta\times 
  T_{rot}={t(R)\over L_\theta}\times T_{rot}
  $$
  \noindent
  where $L_\theta$ is assumed to be well represented by its axisymmetric 
  estimate, i.e., $L_\theta=R\times v_{rot}$. 
  The $\Delta L/L$ curves for NGC~3147 derived from the CO(1--0) and 
  the CO(2--1) data are displayed in
  Figs.~\ref{gastor3147}.

  The gravitational torque maps in Fig.~\ref{torq3147} show 
  that the derived torques change sign following a characteristic 2D 
  {\it butterfly} pattern. The CO contours reveal that for the material in the inner ring
  (radius 2\,kpc, 10\arcsec), part of the gas is trailing the bar, 
  while part of the gas is leading it. The observed gas distribution is representative
  of the time spent by a molecular cloud on a typical orbit at this
  location. The azimuthal average of the torques shown in Fig. \ref{gastor3147}
  suggests that the torques are predominantly negative
  inside a radius of $\sim$2\,kpc, while they become zero or positive outside.

   If the near side is the northeast side, the rotation sense 
   of the galaxy is counterclockwise, and the whole spiral
   structure is trailing with respect to the rotation.
  In the spiral structure winding into the outer $^{12}$CO(1-0) ring, 
  the torques almost cancel out on average but are still predominantly positive.
  Inside the inner ring (r $<$ 2 kpc), however, the dominating torques are negative,
  although weak in absolute value.
  The $^{12}$CO(2-1) emission is more clumpy, and our map suffers from lower sensitivity;
  there is relatively more emission in the central part.
  Although the emission is spread over three quadrants in the center, the negative
  torques are still predominant there (Fig.~\ref{torq3147_2}).
  The negative torques there are due to the central
  asymmetric CO distribution, mainly the nuclear component,
  at radii inside 3-4\arcsec. The component is resolved in $^{12}$CO(2-1),
and is not aligned with the nuclear bar, but rather shifted to the leading
  side; therefore, negative torques dominate.

 \section{Discussion \label{sec:discussion}}
 
 We find that the two molecular structures (inner ring and outer spiral), 
 in addition to a central peak of $^{12}$CO(2-1) emission, dominate the 
 CO map of NGC~3147's inner region. 
 The inner ring coincides with the ring in the dust-only 8\,$\mu$m image,
 and the lopsided central peak falls at the same position as
 the asymmetric nuclear structure
 in the $J-K$ map.
 These features are not present in the stars (e.g., $J$ band, IRAC 3.6\,$\mu$m,
 and \textit{GALEX}), but are rather features of the ISM.
 On the other hand, the outer spiral also appears in the stars as shown by the
 \textit{GALEX} image; however, there is an offset between
 them, as the CO emission falls in the interarm region relative to 
 the outer spiral (see Fig. \ref{galex-co}).

 A probable explanation is that the CO outer spiral is not a permanent structure but is transient.
 The gas participates in a spiral wave and forms stars, and then
 the newly formed stars decouple from the gas component.
 The young stars heat only the dust present at their location, while the bulk
 of the gas has moved to another density condensation.  
 The inward migration of the dense gas with respect to the initial resonance
 has for been discussed by, e.g., \citet{regan03}.
 This phenomenon, because it is very transient, is  statistically rarely encountered.
 In this scenario, the CO clouds would not trace stars of the same age as those 
 that have heated the dust observable at 8\,$\mu$m or of those whose emission 
 is detected in the FUV, but rather they will be the birth site of future stellar generations.

\begin{figure}
  \includegraphics[angle=-90,width=\columnwidth]{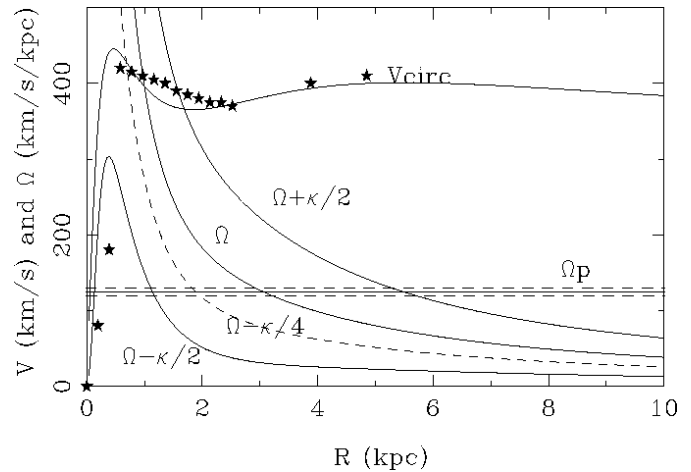}
  \caption{Model rotation curve and derived frequencies $\Omega$,
  $\Omega-\kappa/2$, and $\Omega+\kappa/2$, for NGC~3147. The star symbols
  are the data derived from the CO kinematics, deprojected
  with an inclination of $i$ = 29.5$^\circ$.
  The horizontal line corresponds to 
  $\Omega_p\,=\,125$ km s$^{-1}$ kpc$^{-1}$.}
  \label{vcur}
  \end{figure}

 For the torques computation, we interpret the 
 results by identifying the resonances with the bar.
 Let us note that CO resonant rings have already been observed
 in weakly barred galaxies, where the bar is inconspicuous in the optical
\citep[e.g., NGC~5005, NGC~7217:][]{sakamoto00,combes04}.
 The rotation curve derived from the CO kinematics
 is rather peaked in the center, and corresponds well
 to the gravitational model obtained from the NIR
 image.  Both are fitted by an axisymmetric mass
 model, with a rotation curve as shown in Fig.~ \ref{vcur},
 which allows us to derive the characteristic dynamical frequencies.

 If the stellar bar ends at or just inside its ultra-harmonic resonance
 (UHR), as is canonical for bar dynamics, this means
 that the pattern speed of the bar would be roughly
 $\Omega_p\,=\,120-130$ km s$^{-1}$ kpc$^{-1}$, corresponding to a bar
 with one or two inner Lindblad resonances, as is common.
 The presence of these ILRs is not certain, however, given
 the possible non-circular motions in the bar. 
 The inner CO ring would then correspond to the UHR
 (r\,=\,2\,kpc, 10\arcsec) just inside corotation, which we estimated to 
 be at a radius of $\sim$15\arcsec\ (3\,kpc) in Sect. \ref{sec:kinematics}.
 The measurements of the inner ring radius and bar length suggest that
 the bar/oval ends roughly at the onset of the inner ring.  However, 
 the considerable uncertainties of these measurements and the finite 
 width of the ring lead to an uncertainty in the pattern speed $\Omega_p$ 
 of the bar.

 The outer CO spiral corresponds to the spiral
 structure emerging from the bar, beyond its
 corotation.
 The pitch angle of these spiral arms is quite small, 
 as is expected for a weak bar.
 In this scenario, the gas in the inner CO ring is inside
 corotation; due to the negative torques there, the gas
 loses angular momentum and is therefore 
 flowing inwards.
 Due to the weak bar, the rate of inflow
 is relatively low, with a time-scale
 of about 5 rotations (Fig. \ref{gastor3147}).

 The amount of gas flowing inwards is consistent with
 what might be needed to fuel the Seyfert~2 nucleus.
 The gas surface density in the central kiloparsec is 
 about 10~M$_{\odot}$\,pc$^{-2}$; 
 if the gas at 1\,kpc from the center falls in within 5 dynamical time-scales 
 (of $\sim$20 Myr),
 then the feeding rate is about 0.5 M$_\odot$\,yr$^{-1}$.
 Assuming a mass-to-energy conversion efficiency
 of $\epsilon\sim10$\% ($L = dM/dt c^2 \epsilon$), then 
 a luminosity of order 3 $\times$10$^{45}$ erg s$^{-1}$ is accounted
 for, more than enough for this low-luminosity AGN.
 
 The spatial resolution of the CO maps is not
 yet sufficient to determine more precisely 
 whether the gas is actually feeding the AGN, but we
 can at least conclude that we are observing the gas inflow
 into the $\sim$100--200 pc scale circumnuclear region. 

 \section{Conclusions}
 \label{sec:Conclusions}
 The molecular gas in the Seyfert 2 galaxy NGC~3147 has been 
 mapped with high resolution (1\farcs9 $\times$ 1\farcs6 Gaussian beam
 for the $^{12}$CO(1-0) line) inside a radius of 25$^{\prime\prime}$ ($\sim$5 kpc).
 The CO emission shows a central peak (mainly in $^{12}$CO(2-1))
 and a ring at distance of 2\,kpc.
 In $^{12}$CO(1-0) also an outer spiral at distance of 
 4\,kpc is detected.
 The observed CO has a mean line intensity
 ratio $I_{21}/ I_{10}$  $\sim$0.7, consistent with the optically 
 thick emission expected in the nuclei of spiral galaxies, 
 and broadly regular kinematics with some evidence for local
 non-circular motions.

 Comparing the molecular gas distribution with tracers of  
 star formation, we find that central emission and the inner 
 CO ring coincide well with the FUV (\textit{GALEX}) and 8\,$\mu$m 
 (\textit{Spitzer}) emission, while the outer CO spiral structure/ring is 
 instead located
 within an interarm interval 
 detected by \textit{GALEX} and \textit{Spitzer}. 
 We interpret this partial disagreement as due to the transient 
 nature of the outer CO ring.
 It would be a structure 
 in re-condensation that traces a future stellar generation, but 
 is decoupled from those that have heated the dust observable at 
 8\,$\mu$m or from whose emission is detected in the FUV.

With a NIR image obtained with the Canada-France-Hawaii Telescope,
we have identified the presence of a weak bar in NGC~3147, 
a galaxy classified as non-barred in the optical. 
This stellar bar acting on the gas produces gravity  torques.
 A portion of the gas present 
 in the inner $^{12}$CO(1-0) ring is trailing the bar, while part 
 of the gas leads it. The gravity torques are negative inside 
 a radius of $\sim$2\,kpc, while they become zero or positive outside. 
 In the outer $^{12}$CO(1-0) spiral the torques almost cancel 
 out on average, but are positive. In the more clumpy inner $^{12}$CO(2-1) ring, 
 the predominant torques are negative. 
 We interpret these results by identifying the resonances with the bar. 
 The inner CO ring would correspond to the ultra-harmonic resonance (r
 = 2 kpc) just inside corotation, and the outer CO spiral
 to the spiral structure emerging from the bar, beyond
its corotation, which we estimate is located at $\sim$3\,kpc.
 In the inner CO ring, the gas is inside corotation due to the negative 
 torques there; it loses angular momentum and is flowing inwards.

 NGC~3147 is not the first case in the NUGA sample where inflowing gas has been found: 
 NGC 2782 \citep{hunt}, and NGC 6574 \citep{lindt}
 also show this feature.   However, only NGC 2782 has a strong
 and significant inflow, due to a nuclear bar
 embedded in the primary one and to gas aligned with the nuclear bar.
 The amount of inflowing gas in NGC~3147, quantified
 from the gravity torques computation, is modest but could still be sufficient to feed the
 Seyfert 2 nucleus at a rate of about 0.5 M$_\odot$\,yr$^{-1}$. 
 Although higher spatial resolution is needed to determine whether the 
 gas is actually feeding the AGN, we are observing the gas inflow to the 
$\sim$100--200 pc central region.

  \begin{acknowledgements}
  The authors would like to thank the anonymous referee, 
  whose comments have been very instructive 
  and useful for improving the original version of the paper.
  We thank the scientific and technical staff at IRAM for their 
  work in making our 30\,m and PdBI observations possible.
  V. Casasola is pleased to acknowledge the hospitality and 
  stimulating environment provided by the Observatoire de 
  Paris-LERMA, where part of the work on this paper was done 
  during her stay in Paris, thanks to the EARA agreement. 
  In this work, we have made use of \textit{MAST/GALEX} 
  images and the \textit{Spitzer} archive.
  \end{acknowledgements}

\end{document}